\documentclass[aps,onecolumn,nobibnotes,nofootinbib,preprint,superscriptaddress,showpacs]{revtex4}
\usepackage{graphicx}
\usepackage{amssymb}
\usepackage{amsmath}
\usepackage{amsfonts}
\usepackage{latexsym}
\usepackage{slashed}
\usepackage{multirow}
\newcommand{\be}{\begin{eqnarray}}
\newcommand{\ee}{\end{eqnarray}}

\newcommand{\gev}{\rm \, GeV}

\def\be{\begin{eqnarray}}
\def\ee{\end{eqnarray}}
\newcommand{\nn}{\nonumber}
\def\ltap{\ \raise.3ex\hbox{$<$\kern-.75em\lower1ex\hbox{$\sim$}}\ }
\def\gtap{\ \raise.3ex\hbox{$>$\kern-.75em\lower1ex\hbox{$\sim$}}\ }

\newcommand{\bv}{\ensuremath \not \!\! B }
\newcommand{\lv}{\ensuremath \not \!\! L }
\newcommand {\unit} [1] {\; \mathrm {#1}}

\begin{document}
\author{Joshua T. Ruderman}
\email{ruderman@berkeley.edu}
\affiliation{Department of Physics, University of California, Berkeley, CA 94720}
\affiliation{Theoretical Physics Group, Lawrence Berkeley National Laboratory, Berkeley, CA 94720}

\title{A Collective Breaking of R-Parity}
\author{Tracy R. Slatyer}
\email{tslatyer@ias.edu}
\affiliation{School of Natural Sciences, Institute for Advanced Study, Princeton, NJ 08540, USA}

\author{Neal Weiner}
\email{neal.weiner@nyu.edu}
\affiliation{Center for Cosmology and Particle Physics, Department of Physics, New York University, New York, NY 10003, USA}
\affiliation{School of Natural Sciences, Institute for Advanced Study, Princeton, NJ 08540, USA}


\begin{abstract}
Supersymmetric theories with an R-parity generally yield a striking missing energy signature, with cascade decays concluding in a neutralino that escapes the detector. In theories where R-parity is broken the missing energy is replaced with additional jets or leptons, often making traditional search strategies ineffective. Such R-parity violation is very constrained, however, by resulting $B$ and $L$ violating signals, requiring couplings so small that LSPs will decay outside the detector in all but a few scenarios. In theories with additional matter fields, R-parity can be broken {\em collectively}, such that R-parity is not broken by any single coupling, but only by an ensemble of couplings. Cascade decays can proceed normally, with each step only sensitive to one or two couplings at a time, but $B$ and $L$ violation requires the full set, yielding a highly suppressed constraint. $s$-channel production of new scalar states, typically small for standard RPV, can be large when RPV is broken collectively. While missing energy is absent, making these models difficult to discover by traditional SUSY searches, they produce complicated many object resonances (MORes), with many different possible numbers of jets and leptons. We outline a simple model and discuss its discoverability at the LHC.
\end{abstract}


\maketitle

\section{Introduction}
Supersymmetry has long been an exciting solution to the hierarchy problem, both for the elegance of the concept and its dramatic experimental implications. Squarks and gluinos have strong production processes, and many superpartners may appear in the subsequent cascades. 

A common element in many searches for supersymmetry is missing energy (MET). Possible operators $udd$, $qld$, $lle$ and $lh$ all lead to various signals of lepton- or baryon-number violation, and are thus strongly constrained, in some cases requiring couplings of $10^{-8}$ or smaller. Rather than tolerate such exponentially small values, these operators can be easily forbidden under the assumption of {\em R-parity}, under which superpartners change sign relative to their Standard Model counterparts (formally, with a parity $P_R = 2S + 3B-L$).  The lightest R-parity odd particle is consequently stable, and thus must be neutral (for cosmological constraints). As a result, cascade decays of SUSY particles naturally conclude with two invisible particles, making MET a robust prediction of a wide range of SUSY models.

The principle challenge for SUSY in light of this argument is that, simply put, such signals have not been seen~\cite{ATLASSUSY,CMSSUSY}.

No search has yet shown a robust sign of SUSY with its associated missing energy signal, despite the preferred parameter space coming under increasing tension from a wide range of searches at LEP, the Tevatron, and (in particular) the LHC. One possibility is that, of course, the particles are Just Around The Corner, at masses that have yet to be robustly tested by the LHC, but if we are to continue to entertain the idea of supersymmetry, we must confront our prejudice that MET is an inevitable signature of the framework. If MET is absent, SUSY may be present already in the data, but difficult to extract against large QCD backgrounds.

The suppression of MET is possible even with a stable LSP, but this only occurs in particular classes of models such as Stealth SUSY~\cite{Fan:2011yu, Fan:2012jf}, compressed SUSY~\cite{LeCompte:2011fh, Murayama:2012jh}, or models with lengthened cascades~\cite{Baryakhtar:2012rz}.  A more obvious solution is to abandon R-parity~\cite{Hall:1983id}, allowing the LSP to decay, removing new sources of MET from high energy processes. 

The problem with this approach is that including R-parity violation (RPV) operators in the theory reintroduces the dangerous B and L violation that it was included to prohibit. In the presence of generic RPV couplings, limits on \bv and \lv processes (for reviews that discuss limits on RPV see~\cite{Dreiner:1997uz, Barbier:2004ez}) require terms that typically cause the LSP to decay outside the detector, meaning that MET -- and the strong limits on SUSY -- remain. If one considers {\em only} the $udd$ operator, with a flavor-generic coefficient that saturates the maximum value allowed by $n-\bar n$ oscillations ($\sim10^{-8}$), 
then a squark LSP, which decays to two jets, will have a decay length of $10$'s of cm leading to decays within the detector and suppressed missing energy (however note that with a decay length this long many decays will still occur outside the calorimeter, leading to a tail of events with significant MET).  If the LSP is not a squark, the decay is 3(-or-more)-body and outside the detector.  For recent phenomenological studies of hidden SUSY with RPV see Refs.~\cite{Graham:2012th, Brust:2012uf}.

One possible resolution to this tension between indirect limits and keeping prompt decays is to assume that the RPV operators have non-generic flavor structures.  For example, if one imposes Minimal Flavor Violation~\cite{D'Ambrosio:2002ex} on the coefficient of $udd$, baryon number violating constraints, which involve first generation quarks, are suppressed while prompt decays can proceed through heavier generations~\cite{Nikolidakis:2007fc, Smith:2008ju, Csaki:2011ge}.  We instead pursue a novel and orthogonal implementation of RPV where generic flavor couplings are allowed. 

 If we extend the MSSM with additional fields, a new possibility presents itself: namely, that the RPV occurs {\em collectively}. When looking only at a few couplings, a consistent R-parity assignment can be made for the new fields, but when considering the full ensemble, no consistent assignment of R-parities is possible.  We call this class of models Collective R-Parity Violation (CRPV). This technique for symmetry breaking was first introduced in the context of the hierarchy problem \cite{ArkaniHamed:2001nc, ArkaniHamed:2002qx, ArkaniHamed:2002qy}. The important consequence is that diagrams of symmetry-breaking effects must involve all of the couplings, and often no tree-level process can occur, or loop processes are pushed to higher order.  

In contrast, cascade decays sample only one (or two) operators at a time, and consequently, decays can be prompt inside a detector. This offers an interesting prospect - the possibility of flavor-generic/anarchic RPV couplings, and prompt decays at the LHC of SUSY events into jets, while retaining consistency with other observations.

Any implementation of supersymmetry must of course address the implications of the  discovery of what is probably the Higgs boson with $m_h \approx 125$~GeV~\cite{ATLASHiggs, CMSHiggs}.  
The observed Higgs mass is significantly heavier than the tree-level upper bound in the MSSM, $m_h \le m_Z$.
This leaves two options: (1) the Higgs mass is determined by radiative corrections from the top and stop loop, in which case one requires $m_{\tilde t} \gtrsim 1$~TeV and the theory is fine-tuned at the $\sim 1\%$ level or worse~\cite{Hall:2011aa}, or (2) new physics beyond the MSSM raises the Higgs mass, such as interactions between the Higgs and a singlet~\cite{Hall:2011aa} or the presence of non-decoupling $D$-terms~\cite{Batra:2003nj, Maloney:2004rc}. Natural electroweak symmetry breaking motivates option~(2), in which case the sparticle masses are unconstrained by the Higgs mass and motivated by naturalness to be light and hidden, such as due to CRPV\@.  The new states that lead to CRPV may be unrelated to the physics that raises the Higgs mass, or, more economically, the same sector of new states may raise the Higgs mass and lead to RPV~\cite{sister1, sister2}.

In this paper, we will explore the consequences of the simple idea that $R$-parity is broken collectively. We will begin in section~\ref{sec:pheno} by providing an explicit example of such a model. We discuss its effect on SUSY cascades, as well as flavor constraints, baryon number violation (specifically $\bar n - n$ oscillations and, in the context of gauge mediation, $p^+ \rightarrow \tilde g K^+$), and cosmology. In section \ref{sec:signatures}, we consider the novel multi-jet resonances that can appear in these models.  We discuss both the present limits and new signatures, such as four-jet resonances.  We present additional example models in section~\ref{sec:examples}, in order to emphasize that CRPV is a quite general phenomenon.  Finally, in section~\ref{sec:conclusions}, we give our conclusions.

\section{The Phenomenology of CRPV}
\label{sec:pheno}

\subsection{A Simple Example}
\label{subsec:uddexample}

To illustrate the phenomenology, we work with a simple model of CRPV\@.  We will use this as a template model for the remainder of this paper to demonstrate how the constraints are satisfied and new phenomena are possible. (We include a brief description of some additional example models in section~\ref{sec:examples}.) We add a new set of fields $D, U$ and $\bar D, \bar U$ to the superfields present in the MSSM, $q,u,d,l,e$.\footnote{Here, and for the rest of the paper, we use lowercase letters to denote MSSM fields and capital letters to denote the new fields that communicate the collective breaking of R-parity.}  $U$ and $D$ are vector-like quarks, taken to have the same $\mathbf{SU(3)}_C \times \mathbf{SU(2)}_W \times \mathbf{U(1)}_Y $ quantum numbers as $u$ and $d$, $(\mathbf{\bar 3}, \mathbf{1}, {-2/3})$ and  $(\mathbf{\bar 3}, \mathbf{1}, {1/3})$, while $\bar U$ and $\bar D$ live in the conjugate representations.   We add to the MSSM the additional superpotential terms\footnote{In the interest of compact notation we have dropped SM flavor indices, for example $\lambda_D^{ij} u^i d^j D$.} 		
\be
W &\supset& m_D D \bar D + m_U U \bar U \nn \\
&+& \lambda_{D} \, u d D +  \lambda_{UD} \, UdD + \lambda_{U} \, Udd,
\label{eq:crpvmodel}
\ee
where $m_D$ and $m_U$ are weak scale masses which may be generated by the same physics that generates the $\mu$-term, $\mu \, h_u h_d$.  Note that if any of the $\lambda$'s are zero, we can assign a consistent R-parity to the new fields, but with all together, the theory breaks R-parity.  

Once we add soft-supersymmetry breaking, the above theory is augmented by the usual soft terms of the MSSM and, in general, $A$-terms that complement the $\lambda$ couplings ($a_D \, u d D$, etc.).  Supersymmetry breaking will also generate soft masses for the vector-like quarks which split the scalar and fermionic masses, $V \supset \tilde m_D |D|^2+ \tilde m_{\bar D} |\bar D|^2 + b_D D \bar D + \mathrm{h.c.}$, and similarly for $U$ and $\bar U$.

One may wonder why the usual RPV terms, $u d d, q  ld, l l e$ are not present in the theory of equation~\ref{eq:crpvmodel}.  It is important to note that because of the non-renormalization theorems of supersymmetry, it is automatically technically natural for these operators not to be present and for R-parity to be broken collectively.  Our usual intuition as to why all terms not forbidden by a symmetry are present in an effective theory is based on the presence of corrections that are potentially absent in SUSY theories. Thus, even if we are ignorant as to {\em why} the manifestation of RPV is collective, its appearance in this fashion is completely natural.

\begin{figure}[!h]
\begin{center} 
\includegraphics[scale=0.35]{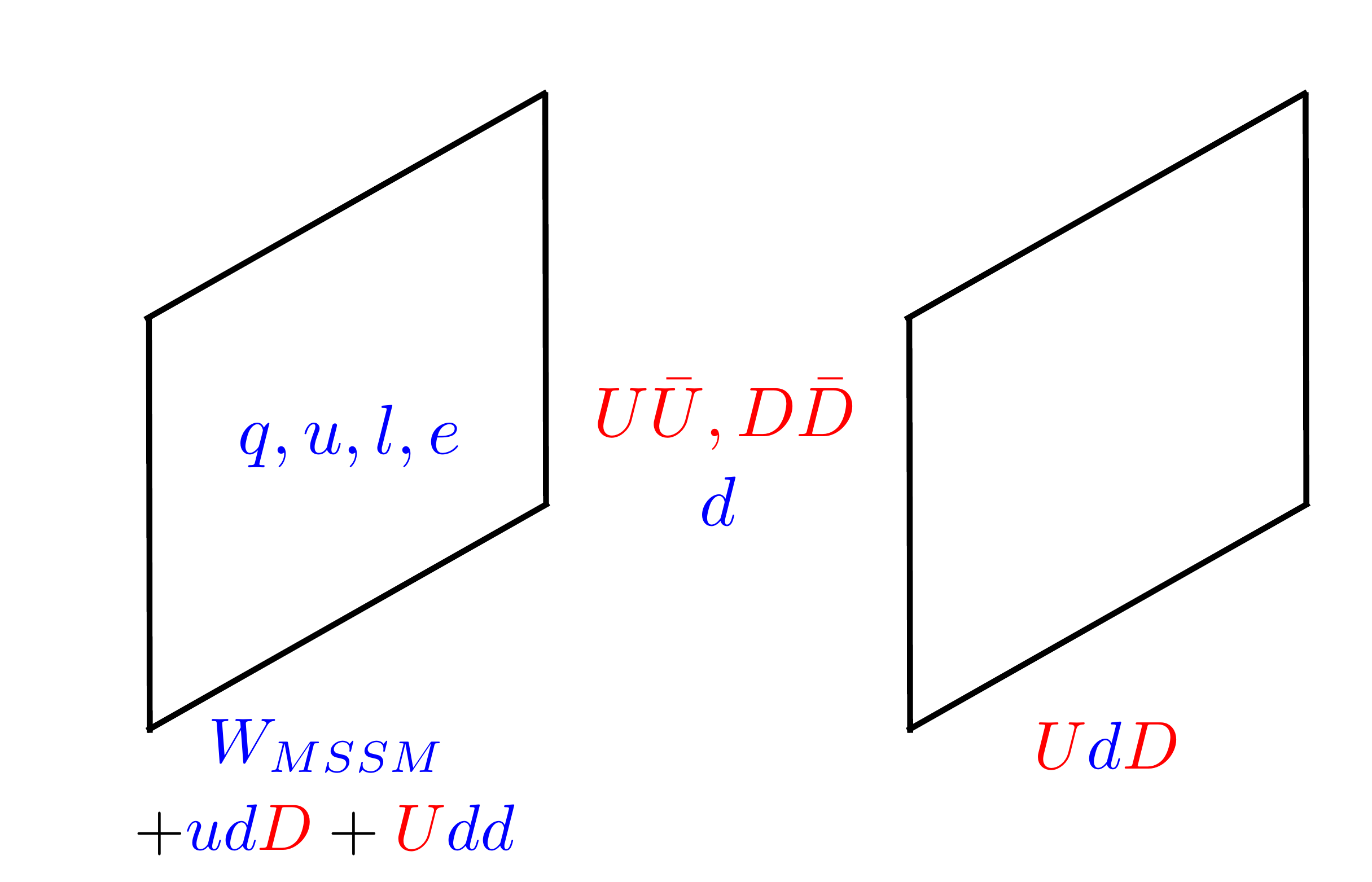} 
\end{center}
\caption{An extra dimensional realization of collective R-parity violation.  The interactions of each brane separately conserve $R$-parity but there is no consistent $R$-parity assignment for both branes.  Therefore, $R$-parity is preserved by the local 5D interactions but broken globally.}
\label{fig:branes}
\end{figure}

That said, it is simple to envision how such a model could be realized from microphysics. One of the simplest ways is to invoke some form of sequestering, such as in an extra dimension.  Then, R-parity may be preserved by the local 5D interactions but broken globally, as illustrated in figure~\ref{fig:branes}.     

We assume that most of the MSSM fields are localized on one brane, specifically $ qule$, but imagine that $U, \bar U, D, \bar D$ and $d$ propagate in the bulk.  On this brane, we include the terms $ U d d$ and $u D d$, which are consistent with R-parity if $U$ and $D$ are neutral.  On the second brane, we add the operator $U D d$ (the operator $U DD$ vanishing identically), which would be consistent with $R$-parity if either $U$ or $D$ is odd and the other one is neutral.  With this setup, the two branes separately preserve $R$-parity but there is no consistent $R$-parity assignment for both branes: $R$-parity is preserved by local interactions but violated by the global geometry.  The low energy theory will then just be the model in eq \ref{eq:crpvmodel} above.    Note that we have included all operators on the MSSM brane consistent with gauge symmetry, R-parity and lepton number: unwanted mass mixings between the MSSM fields and $U$ and $D$ are forbidden by R-parity, such as the superpotential terms $q h_u U, q h_d D, d \bar D$.

The point of this realization is not to claim that it is the simplest or most natural UV completion, but just to illustrate that the form of RPV operators in the IR are highly dependent on the structure of the UV physics, and, as always in supersymmetric theories, symmetries may be broken in the superpotential without the inclusion of the most generic set of operators.

\begin{figure}[!h]
\begin{center} 
\includegraphics[scale=0.65]{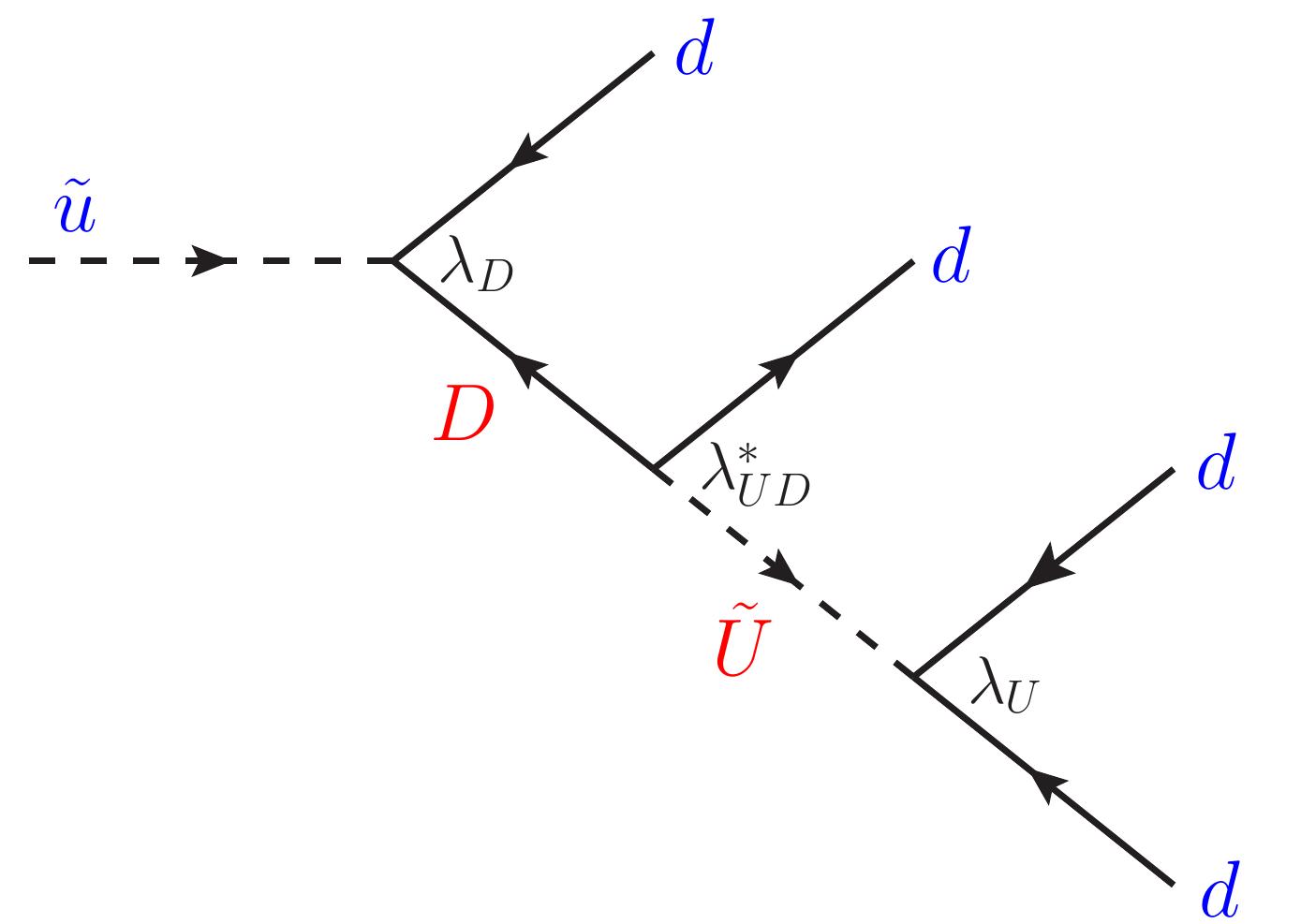} 
\end{center}
\caption{Cascade decays in CRPV.}
\label{fig:cdecay}
\end{figure}

Like usual hadronic RPV (i.e. including $udd$), a generic cascade in this model will lead to a final state with jets.  The primary difference is that the collective breaking leads to higher multiplicity final states. This is because any decay that violates $R$-parity must use all three interactions in equation~\ref{eq:crpvmodel}, necessarily leading to extra jets in the final state.  With usual hadronic RPV, a squark can decay directly into two jets.  This is to be contrasted with figure~\ref{fig:cdecay}, which shows an example decay of a squark with CRPV, which leads to four jets in the final state.  In this model, a gluino or neutralino always decays to 5 jets, compared to 3 jets in conventional RPV.  As long as the $U$ and $D$ states are light, the decays in figure~\ref{fig:cdecay} take the form of sequential 2-body decays.  In this situation, only one coupling is probed at a time when determining whether or not the cascade is prompt.  If the cascade is prompt, MET is removed.

What are the leading constraints on this model?  Both the MSSM sparticles and the $U$ and $D$ states are produced in colliders.  There are constraints on dijet and trijet resonances, as we will discuss in detain in section~\ref{sec:signatures}, however these constraints are weak due to large QCD backgrounds.  The other constraint one can worry about is that from vectorlike fourth generation searches. However, the decays will come not through mass-mixing (and thus not through $W$ or $Z$ emission), but through the cRPV operators, hence, the usual fourth generation searches will be insensitive, with essentially all new colored production resulting in multijet signals. 

Although the collider limits are weak, there are significant indirect constraints, which are the subject of the rest of this section.  We will discuss baryon number violating constraints, which, because of the collectivity, constrain the product $\lambda_D \lambda_{UD} \lambda_U$, in section~\ref{subsec:baryon}.  The individual couplings lead to hadronic flavor violation, as will discuss in section~\ref{subsec:flavor}.  As we will see, the baryon number and flavor violating constraints are both satisfied when each of the couplings satifies $\lambda \lesssim 10^{-2}$, easily allowing for prompt decays even with anarchic flavor couplings.  Baryon number violation is also constrained cosmologically by the requirement that a large enough baryon asymmetry is generated: we discuss the cosmology of CRPV in section~\ref{subsec:baryon}.

\begin{table}[h!]
\begin{center}
\begin{tabular}{|c|c|c|}
\hline
process & leading constraints & limit \\
\hline
\hline
$\Delta B = 2$ &  $N \leftrightarrow \bar N$, double nucleon decay & $\lambda_D \lambda_{UD} \lambda_U \lesssim 10^{-7}$ \\
\hline
$\Delta B = 1$ & $p \rightarrow \tilde G K^+$ & $\lambda_D \lambda_{UD} \lambda_U \lesssim 10^{-16} \left(m_{3/2} \,  / \, \unit{eV}\right)$ \\
\hline
$\Delta F = 2$ & $K \leftrightarrow \bar K$, $D \leftrightarrow \bar D$ & $\lambda_D, \lambda_{UD}, \lambda_U \lesssim 10^{-2}$ \\
\hline
\end{tabular} \end{center}
\caption{\label{tab:constraints}
The leading indirect constraints on the couplings of the model of eq.~\ref{eq:crpvmodel} arising from baryon number violation and flavor violation.  The $\Delta B = 2$ and $\Delta F = 2$ constraints apply generally, while the $\Delta B = 1$ constraint only applies to models with low-scale supersymmetry breaking, where $m_{\tilde G} < m_p$.  For simplicity, we have suppressed flavor indices and the above limits should be viewed as ``worst case" (in general the strongest limits are on the couplings to first and second generation quarks). 
}
 \end{table}

\begin{figure}[!h]
\begin{center} 
\includegraphics[scale=0.75]{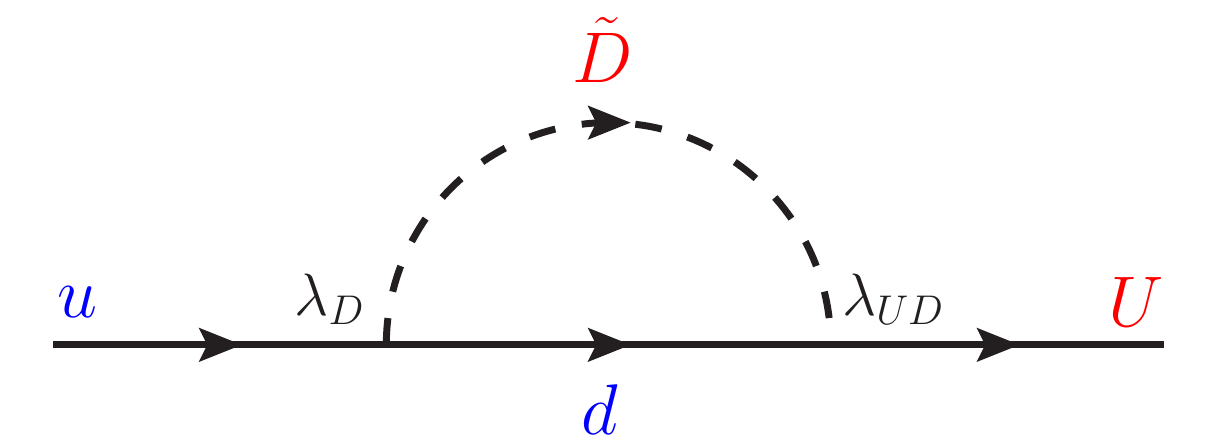} 
\end{center}
\caption{Diagram generating the kinetic mixing of $u$ and $U$.}
\label{fig:kinmix}
\end{figure}

Before proceeding to discuss the constraints in detail, we note that while the collective nature of the RPV is protected by non-renormalization theorems in the superpotential, it is not in the K\"ahler potential. Consequently, diagrams such as figure \ref{fig:kinmix} will generate operators such as 
\be \label{eq:KM}
\frac{\lambda_D \lambda_{UD}}{16 \pi^2} \log(\Lambda/{\rm TeV}) U^\dagger u+ \frac{\lambda_U \lambda_{UD}}{16 \pi^2} \log(\Lambda/{\rm TeV}) D^\dagger d + {\rm h.c.}
\ee
Here, $\Lambda$ represents the scale from which these operators are generated (some UV scale, perhaps a compactification scale). 

The effects of this mixing can be thought of in two ways. At the weak scale we can shift the new fields, e.g. $U\rightarrow U + \epsilon \, u$, where $\epsilon$ is the coefficient of the kinetic mixing in equation~\ref{eq:KM}, and the operator $\lambda_U Udd$ will be mapped into the conventional RPV operator,
\be
\frac{\lambda_U \lambda_D \lambda_{UD}}{16 \pi^2}  \log(\Lambda/{\rm TeV}) udd.
\ee
On the other hand, this will also generate off-diagonal mass terms $\bar D d+\bar U d$. A more natural mapping is to shift $u \rightarrow u + \epsilon \, U$ which will {\em not} generate the usual $udd$ operator, but will introduce off-diagonal Yukawa couplings such as $q D h_d$. Additionally, it will introduce off-diagonal scalar masses; at any scale where SUSY is broken, additional mixings in the scalar mass matrix are also generated radiatively, from diagrams involving the (diagonal) scalar soft masses and A-terms. When coupled with e.g., $Udd$, these mass mixings will lead to a similar phenomenology to $udd$, because after diagonalizing the scalar mass matrix, RPV Yukawa couplings are generated, such as $\tilde u \psi_d \psi_d$.

It is important to note that the collider phenomenology is not dominated by these effects when the $U,D$ fields are present at low energy.  This is because the kinetic/mass mixing induced effects are determined by a product of at least two couplings, $\mathcal{O}(\lambda^2)$, while decays involving $U$ and $D$ follow directly from the couplings in equation~\ref{eq:crpvmodel} at $\mathcal{O}(\lambda)$.

\subsection{Collective Baryon Number Violation}
\label{subsec:baryon}

\begin{figure}[!h]
\begin{center} 
\includegraphics[scale=0.35]{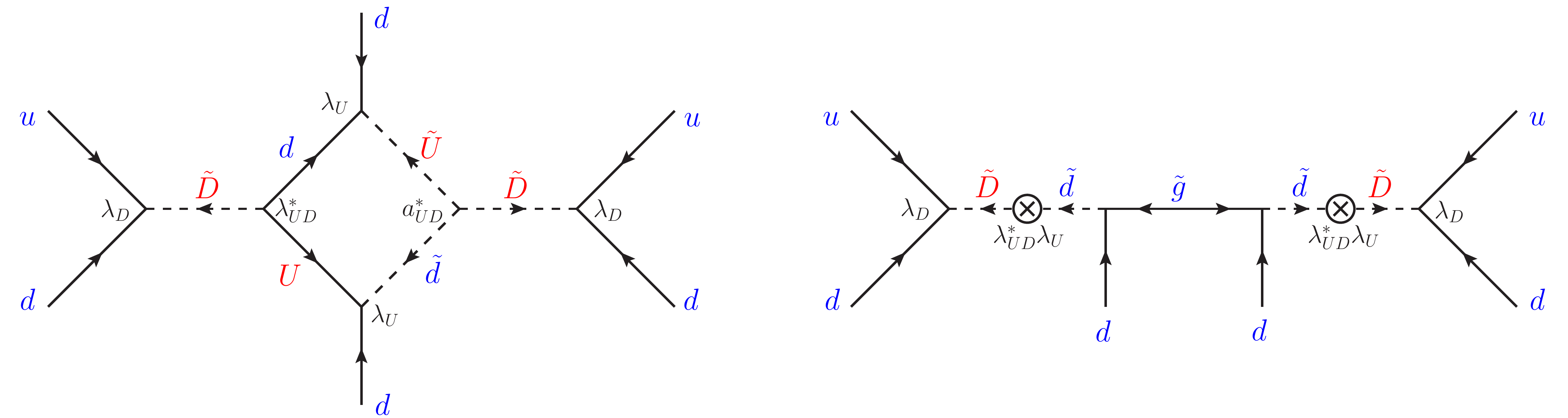} 
\end{center}
\caption{The leading contribution to $n-\bar n$ oscillations and double nucleon decay, $p p \rightarrow K^+ K^+$.}
\label{fig:NNbar}
\end{figure}

As usual in RPV, one cannot break R-parity without breaking either baryon number or lepton number. This theory is no exception. However, just as RPV is collective, so, too, is baryon number violation. Consequently, the diagrams that contribute to e.g., $n -\bar n$ oscillations must involve all the couplings.  Indeed, a simple spurion argument shows that any matrix element that violates baryon number must be proportional to the following product of couplings,

\be \label{eq:DeltaBscaling}
i M_{\Delta B} \propto \left( \lambda_D \lambda_{UD}^* \lambda_D \right)^{\Delta B},
\ee
where, once we include supersymmetry breaking, it is understood that any of the above couplings can be swapped with its respective $A$-term, for example $\lambda_{UD} \leftrightarrow a_{UD}$.
Therefore, $n \bar n$ oscillations, which are $\Delta B = 2$, must include all of the operators {\em twice} (i.e., $O_{\Delta B = 2} \propto (\lambda_{D}  \lambda_{UD}^*    \lambda_{U})^2$), and we show the leading diagrams in figure \ref{fig:NNbar}.  The main constraints actually come from two different processes, $n \leftrightarrow \bar n$, where all external quarks are first generation, and double nucleon decay, $p + p \rightarrow K^+ + K^+$ and $n + n \rightarrow K^0 + K^0$, where two of the external down-type quarks are strange.

Note that both of the diagrams in figure~\ref{fig:NNbar} require supersymmetry breaking: the diagram on the left is proportional to the $A$-term $a_{UD}^* U^* d^* D^*$ and the diagram on the right contains a gluino mass insertion, $M_3^*$.  This is because in the supersymmetric limit $n \bar n$ oscillations and double nucleon decay are protected by the unbroken $R$-symmetry.  Under the $R$-symmetry, the $\Delta B = 2$ operator $(\psi_u \psi_d \psi_d)^2$, has $R$-charge $-2$.  We can treat $a_{UD}^*$ and $M_3^*$ as spurions with charge $+2$, and the allowed operators include $a_{UD}^* (\psi_u \psi_d \psi_d)^2$ and $M_3^* (\psi_u \psi_d \psi_d)^2$.

For $n \bar n$, the matrix element of the diagram on the left is estimated to be,
\be \label{eq:NNbarEstimate}
i M_{n \bar n} \sim \frac{(\lambda_D^{11})^2 (\lambda_{UD}^*)^i \lambda_U^{i1} \lambda_U^{j1}}{(4 \pi)^2 } \frac{(a_{UD}^*)^j}{m_{\tilde D}^4 \bar m^2},
\ee
where $\bar m$ is a mass scale coming from the loop integration, which depends on $m_U, m_{\tilde U}, m_{\tilde d}$.  Here, the flavor indices, $i, j$ are summed over, and  $i,j \neq 1$ because $\lambda_U^{11} =0$.  For double nucleon decay, there are several flavor combinations  of couplings that are constrained, depending on which external squarks are taken to be strange.  For example, taking the central two down-type squarks to be strange, the contribution to the amplitude is given by,
\be \label{eq:NNbarEstimate2}
i M_{NN\rightarrow KK} \sim \frac{(\lambda_D^{11})^2 (\lambda_{UD}^*)^i \lambda_U^{i2} \lambda_U^{j2}}{(4 \pi)^2 } \frac{(a_{UD}^*)^j}{m_{\tilde D}^4 \bar m^2},
\ee
where $i,j \neq 2$.
Suppressing now the flavor indices, which are understood, both $n \bar n$ and double nucleon decay lead to the approximate limit,
\be
\lambda_{D} \lambda_{UD} \lambda_{U} \lesssim 10^{-7} \left( \frac{m_{\tilde D}}{100\unit{GeV}} \right)^4 \left( \frac{\bar m}{100\unit{GeV}} \right)^2 \left( \frac{100\unit{GeV}}{A_{UD}} \right),
\ee
Here $A_{UD}$ is defined by the relation $a_{UD} = \lambda_{UD} A_{UD}$.  We see that this limit is accommodated if all of the couplings are $O(10^{-(2-3)})$, easily allowing for prompt collider decays even if the $\lambda$ couplings have an anarchic flavor structure.  We have derived the above limit by rescaling the limit from conventional R-parity violation, and we note that this limit is uncertain by as much as several orders of magnitude due to unknown nuclear matrix elements (for more details see \cite{Goity:1994dq, Barbier:2004ez}), but this uncertainty does not affect our qualitative conclusion.

 For the right diagram of figure~\ref{fig:NNbar}, the matrix element is $i M \sim \lambda_D^2 \epsilon_{dD}^2 / \tilde m^5$, where $\epsilon_{dD} \sim \lambda_U \lambda_{UD} / (4 \pi)^2 \log (\Lambda / \tilde m)$, $\tilde m$ is the mass scale of the superpartners, and $\Lambda$ is the scale of supersymmetry breaking.  Here, we have suppressed the flavor indices, but as above all external quarks should be taken to be taken to be first generation for $n \bar n$ oscillations and two external quarks should be strange for double nucleon decay.
The resulting limit is similar to the limit from the first diagram if supersymmetry is broken at a high-scale such that the logarithm is large,
\be
\lambda_{D} \lambda_{UD} \lambda_{U} \lesssim 10^{-7} \left( \frac{\tilde m}{100\unit{GeV}}\right)^4 \left(\frac{10}{\log (\Lambda / \tilde m)} \right).
\ee

\begin{figure}[!h]
\begin{center} 
\includegraphics[scale=0.45]{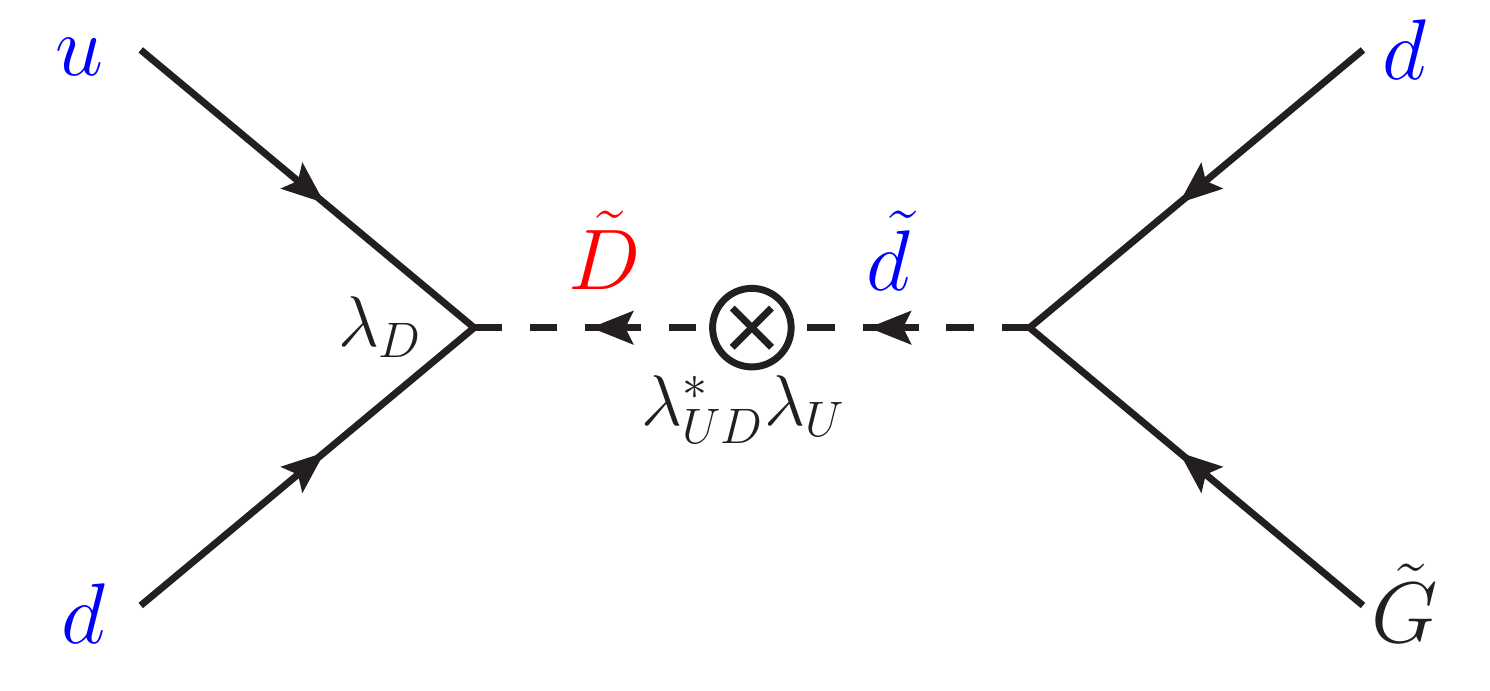} 
\end{center}
\caption{\label{fig:ProtonDecay}
Proton decay.}
\end{figure}

Remarkably, the relative safety (when compared to traditional RPV) of this theory appears even for $\Delta B=1$. For instance, in the case of gauge mediation, one can be concerned about the proton decay into a gravitino $p \rightarrow \tilde G+K^+ (\pi^+)$.  For traditional hadronic RPV with $udd$, the coupling must be smaller than $10^{-15} (m_{3/2} / \unit{eV})$~\cite{Choi:1996nk}, which means that low-scale supersymmetry breaking typically forces hadronic RPV decays to occur well outside of the detector.  For our model, eq~\ref{eq:DeltaBscaling} implies that proton decay to gravitino requires the product $\lambda_D \lambda_{UD}^* \lambda_U$, as shown in figure~\ref{fig:ProtonDecay}.  The matrix element is given by,

\be
i M_{\Delta B = 1} \sim \frac{\lambda_D \lambda_{UD}^* \lambda_U}{(4 \pi)^2} \frac{\log{\Lambda / \tilde m}}{\tilde m^2 F}.
\ee
We have suppressed the flavor indices, but for proton decay to $\pi^+$ ($K^+$), all external fermions are first generation (one external down-type squark is strange).  The resulting limit is,
\be \label{eq:protondecay}
\lambda_D \lambda_{UD}^* \lambda_U \lesssim 10^{-16} \left(\frac{m_{3/2}}{\unit{eV}}\right) \left( \frac{\tilde m}{100\unit{GeV}} \right)^4 \left(\frac{10}{\log (\Lambda / \tilde m)} \right)
\ee
These limits can be satisfied, even for ultra-low mediation scales, by taking the individual couplings $\lambda \lesssim \mathcal{O}(10^{-(5-6)})$, which allows for decays within the detector, even with anarchic flavor couplings for the $\lambda$.  We note that the above constraints are found by applying the limits on $p \rightarrow \nu + K^+$ and $p \rightarrow \nu + \pi^+$.  The strongest limit is on the decay to kaons, $6.7\times10^{32}$ years, and not pions which have a limit of $2.5\times10^{31}$ years~\cite{pdg}.

In summary, the collectivity of CRPV automatically allows a suppression of the dangerous baryon number violating terms. Similarly, in models where instead lepton number is violated, the signals are likewise suppressed. 

\subsection{Flavor Violation}
\label{subsec:flavor}

There are a broad array of constraints on  R-parity violation from precision observables and flavor. Several of these limits, despite being much weaker than constraints from baryon number violating processes for conventional R-parity violation, apply almost unchanged to the collective case. As discussed in section~\ref{subsec:uddexample}, mixing between the $U$ and $u$ and $D$ and $d$ fields can be radiatively generated at one-loop level, but the resulting contributions to flavor violation are equal or smaller to those we will describe in this section.

For real R-parity violating couplings, most of the the strongest constraints arise from new four-fermion operators that contribute to the mixing of neutral mesons, $K \bar K$ and $B \bar B$, which we show in table \ref{tab:flavor}. Additional strong constraints can arise from rare decays of hadronic B mesons. These limits are not suppressed by collectivity, and do not depend on the product of all three couplings, but are generally comparable to those arising from $n -\bar n$ oscillations for anarchic couplings. 

For $K \bar K$ and $B \bar B$ mixing, the same box diagram that dominates in the usual case also contributes in the collective case: simply applying the limit from the conventional case to the collective case, products of pairs of R-parity violating couplings are constrained to satisfy $|\lambda|^2 \lesssim 2-3 \times 10^{-4} (\tilde{m} / \mathrm{100 GeV} )^2$ \cite{Barbier:2004ez} (the combination $|\lambda_{U}^* \lambda_{D}|$ is not directly constrained, but all other pairs of couplings must satisfy this limit; in the non-flavor-anarchic case, the relevant flavor indices are shown in table \ref{tab:flavor}). A slightly stronger limit may be obtained where the mass of the scalars is comparable to the mass of the top quark ($|\lambda^2| \lesssim 6 \times 10^{-4} (\tilde{m} / \mathrm{100 GeV})$), but this does not change our qualitative conclusions. As expected, these limits are comparable to the estimated constraints presented in table \ref{tab:flavor}.

CP violation can be introduced if the R-parity violating couplings are complex. The strongest constraint arises from the neutral $K \bar K$ system. In the conventional case, a competitive contribution to the parameter describing $\Delta S = 1$ direct CP violation arises from a one-loop diagram; in our toy model, this same diagram can be generated by any one of the three couplings, and consequently the usual constraint can be directly translated to $\mathrm{Im} \left(\lambda_{uDd}^2, \, \lambda_{Udd}^2, \, \lambda_{UDd}^2 \right) \lesssim 10^{-8} (\tilde{m}/ \mathrm{100 GeV})^2$ \cite{Barbieri:1985ty,Barbier:2004ez}, assuming flavor-universal couplings. There are also bounds on CP violation from contributions to the neutron electric dipole moment at two-loop level, but these are weaker than the $K \bar K$ limits for generic flavor couplings ($\mathrm{Im} (\lambda^2) \lesssim 10^{-2}$). In general, we will assume real couplings.

Measurements of the hadronic branching ratios of the $Z$, the asymmetry parameters $A^b$ and $A^b_\mathrm{FB}$, and the rare decay $b \rightarrow s \gamma$ place bounds on $R$-parity violation, but even in the conventional case without collectivity, these particular channels only require the coupling $\lambda$ to be $\mathcal{O}(1)$. Rare hadronic decays of B mesons, other than the channel mentioned above, set constraints on products of pairs of couplings at the $|\lambda|^2 \lesssim$ few $\times 10^{-2}-10^{-3}$ level, with the strongest limits coming from $\bar B^0 \rightarrow \pi^0 \bar K^{0*}$, $B^- \rightarrow \pi^0 K^-$.

Altogether, these flavor constraints seem to require couplings in the $10^{-2}$ range, without assuming any particular flavor structure, which still allows us prompt decays, and couplings much larger than what is allowed from traditional RPV.

\begin{table}[h!]
\begin{center}
\begin{tabular}{|c|c|c|c|c|}
\hline
process & operator & scale (TeV)& constrained couplings & approx. limit ($\sqrt{\lambda \lambda'}$)\\
\hline
\hline
$\Delta m_K$ & $(\bar s_R \gamma^\mu d_R)^2$& 980&$ \lambda_D^{i1} \lambda_D^{i2}, \, \lambda_U^{13} \lambda_U^{23}, \, \lambda_{UD}^1 \lambda_{UD}^2$&$0.03$ \\
\hline
$\Delta m_D$ &$(\bar c_R \gamma^\mu u_R)^2$ &1200 & $\lambda_D^{1i}\lambda_D^{2i}$&$0.03$\\
\hline
$\Delta m_{B_d}$&($\bar b_R \gamma^\mu d_R)^2$& 510 &$\lambda_D^{i1} \lambda_D^{i3}, \,  \lambda_U^{12} \lambda_U^{23}, \, \lambda_{UD}^1 \lambda_{UD}^3$&$0.05$  \\
\hline
$\Delta m_{B_s}$&($\bar b_R \gamma^\mu s_R)^2$ &110 & $ \lambda_D^{i2} \lambda_D^{i3}, \, \lambda_U^{12} \lambda_U^{13}, \, \lambda_{UD}^2 \lambda_{UD}^3$&$0.10$ \\
\hline
\end{tabular} \end{center}
\caption{\label{tab:flavor}
The limits on CRPV couplings coming from $\Delta F = 2$ processes.  Our model contributes to the meson mixing processes listed in the first column by generating the $R-R$ 4-Fermi operators listed in the second column.  The limits on the scale suppressing these operators are given in the third column~\cite{Isidori:2010kg} (the Lagrangian terms take the form $\psi \bar{\psi} / \Lambda^2$ where $\Lambda$ is the scale being constrained).  In the fourth column we list the products of couplings in our model that are constrained by each process.  The final column shows the approximate limit on the geometric mean of the relevant couplings assuming that $m_{U, D} \approx 100$~GeV (this limit gets weaker linearly with $m_{U,D}$ as the masses are raised).
}
 \end{table}

\subsection{Cosmology}
\label{subsec:cosmo}

RPV presents a challenge for cosmology because baryon and lepton number violation have the potential to wash out a pre-existing lepton or baryon asymmetry.  CRPV is no exception, although the constraints can be alleviated in CRPV because of the possibility of a lighter gravitino, as  we now discuss.

In the SM and MSSM, the sphalerons preserve $B-L$ and violate $B+L$.  If baryon number, or lepton number, is also violated through RPV interactions, then any pre-existing $B$ or $L$ asymmetry, such as can result from leptogenesis, is washed out completely.  The requirement that the RPV interactions are out of equilibrium at the temperature of the electroweak phase transition leads to the strong constraint $\lambda \lesssim 10^{-7}$ for the $LLE, LQD$, and $UDD$ couplings, typically making RPV irrelevant in colliders.  The same constraint is true for CRPV, and the collectivity does not help.  For example, in the model introduced above, as long as each coupling satisfies $\lambda_D, \lambda_{UD}, \lambda_U \gtrsim 10^{-7}$, all interactions equilibrate and the baryon and lepton asymmetries are washed out.  One obvious exception that applies to both RPV and CRPV is if the baryon asymmetry is generated through electroweak baryogenesis.

A second exception, where  CRPV has a real advantage, relies on the observation that the sphalerons also preserve two $L_i - L_j$ asymmetries.  Therefore, an initial $L_i - L_j$ asymmetry is preserved both by the sphalerons and by any baryon number violating interactions coming from RPV or CRPV.  This lepton flavor asymmetry can be converted to a baryon asymmetry by lepton mass effects and by slepton mass effects if the sleptons are lighter than the temperature where the sphalerons decouple~\cite{Dreiner:1992vm, Davidson:1998za}.  However, the slepton mass matrices generically violate both $L_i - L_j$ asymmetries, removing this option unless the flavor violation in the slepton mass matrix is sufficiently suppressed, $\theta_{12} \lesssim 10^{-4}$, $\theta_{13}, \theta_{23} \lesssim 10^{-5}$ (it is sufficient for two out of three of these inequalities to be satisfied)~\cite{Endo:2009cv}.  These constraints are more stringent than experimental limits on lepton flavor violation and are violated by Planck suppressed operators $m_{3/2}^2 < 10^{-5} m_{\tilde l}^2$.  There is tension, for conventional RPV, between this limit and the requirement that $m_{3/2} > m_p$, to avoid the stringent constraint coming from proton decay to gravitino, which is discussed above.  On the other hand, collectivity protects CRPV from proton decay to gravitino, and we see from equation~\ref{eq:protondecay} that $m_{3/2}$ can be light enough to preserve the $L_i-L_j$ asymmetries, while collider decays remain prompt and the proton remains sufficiently long-lived.

\section{Multi Object Resonances in CRPV}
\label{sec:signatures}

Collective RPV has reduced missing energy, compared to R-parity conserving supersymmetry, and this means that it will be more difficult to discover in colliders.  It is worth considering whether this model contains any novel signatures that can lead to discovery, and indeed there are many.

In CRPV, the lightest superpartner decays, through the ensemble of R-parity violating couplings, into SM final states.   Any squark (or gluino) produced will ultimately end in some multi-particle final state that  reconstructs to a resonance. These multi-object resonances (MORs) should be generally present in any RPV theory.  CRPV has two important differences compared to traditional RPV.  First, the resonances are more generically prompt in CRPV because larger couplings are allowed.  Second, since multiple couplings must be probed to violate R-parity, the final state contains more particles and a richer resonance structure ({\it i.e.} more MORs).

Consider, for example, the model introduced in section~\ref{sec:pheno}.  Many resonances are possible, depending on the state produced.  The scalar $\tilde D$ and $\tilde U$ states decay to dijets through the $\lambda_D$ and $\lambda_U$ couplings, respectively.  On the other-hand, the fermionic $D$ and $U$ states must decay into at least 3 jets each, for example  $D$ can decay into a jet plus $\tilde U$ through the $\lambda_{UD}$ coupling, and $\tilde U$ can subsequently decay into two jets through the $\lambda_U$ coupling.  Moving on to the SM superpartners, an up or down type squark decays into a 4-jet final state, as illustrated in figure~\ref{fig:cdecay}.  This 4-jet final state contains a 3-jet $D$ (or $U$) sub-resonance and a dijet $\tilde U$ (or $\tilde D$) sub-resonance.  This is to be contrasted with conventional RPV where a squark decays directly into two jets.  A gluino, if produced, must decay into at least 5 jets (compared with three in conventional RPV)!  These various multi-jet resonance possibilities are summarized in table~\ref{tab:Njets}.  As we discuss in more detail below, it is possible to re-interpret some existing searches for the hadronic resonances of conventional RPV as limits on our model.   These limits, which are summarized in table~\ref{tab:Njets}, are very mild.

\begin{table}[h!]
\begin{center}
\begin{tabular}{|c||c|c||c|c|}
\hline
 \multirow{2}{*}{particle}  & \multicolumn{2}{c||}{min. jets} &  \multicolumn{2}{c|}{direct CRPV limit}\\
  \cline{2-5} & CRPV & vanilla RPV & limit & search \\
\hline
\hline
$\tilde U, \tilde D$ & 2 & - &86, 81 GeV& ALEPH~\cite{Heister:2002jc}\\
\hline
$U, D$ & 3 & - & 90~GeV & CDF~\cite{Aaltonen:2011sg}\\
\hline
$\tilde u, \tilde d$ & 4 & 2 & - & - \\
\hline
\multirow{2}{*}{$\tilde g, \tilde N_1$} & \multirow{2}{*}{5} &\multirow{2}{*}{3} &$m_{\tilde g} \gtrsim 500$~GeV&\multirow{2}{*}{CMS~\cite{Chatrchyan:2011cj, CMStrijet5}}\\
&&&(when $m_{\tilde D/\tilde U} \sim m_{\tilde g}$) &\\
\hline
\end{tabular} \end{center}
\caption{\label{tab:Njets}
Various multijet resonances of CRPV and the limits on them.  On the left of the table, we compare the number of parton-level jets produced in the decay of various colored states to the number of jets in conventional (vanilla) RPV.   On the right of the table, we show the leading collider limits on the CRPV states, coming from searches for hadronic resonances from LEP, the Tevatron, and the LHC.  These limits are very mild.
}
 \end{table}

Another possibility afforded by R-parity violation is the single production of colored scalars.  Single production cross-sections can be larger in CRPV, compared to conventional RPV, because larger couplings are allowed.
 Below, we will consider the possibility of $s$-channel production of $\tilde D$, which can lead to a detectable 4-jet resonance.

Throughout this section, we focus on the irreducible collider limits on multijet final states.  We note that there are also spectrum dependent signatures that can facilitate a faster discovery.  For example, top quarks may be produced in gluino decays to stops, and if charginos are produced, their decays may produce $W$ bosons which subsequently decay to leptons.  In these situations, same-sign dilepton searches can set powerful constraints, even when missing energy is removed by RPV~\cite{Brust:2012uf}.  We stress that these signatures are model-dependent: for example if the gluino-stop mass splitting is less than $m_t$, tops are squeezed out and gluino decays can pass entirely through light flavor squarks.  Since we are interested in models of hidden SUSY, we choose to focus here on irreducible multijet signatures and we do not consider spectra-dependent multilepton signatures.

\subsection{Dijets in CRPV}

We begin by discussing dijets, which follow from the decay of the scalar components of the vector-like quarks, $\tilde D$ and $\tilde U$.  These scalars can be pair produced at LEP2 through the neutral current and at hadron colliders through QCD.  These scalars can also be singly produced at hadron colliders for large enough values of $\lambda_U$ and $\lambda_D$, however there is no limit when all flavor components satisfy $\lambda_{U, D}\lesssim 0.3$.  In this section, we focus on the limits on the irreducible pair production.  We return to single production below when we discuss 4-jet resonances. 

If the scalars are light enough, then they would have been pair produced at LEP2.  The 4-jet final state resulting from $\tilde U,\tilde D$ production is constrained by searches for squark production with conventional R-parity violation.  The strongest limits are $m_{\tilde D} > 81$~GeV and $m_{\tilde U} > 86$~GeV, at 95\% C.L., from the ALEPH search for RPV~\cite{Heister:2002jc}.  The slightly stronger limit on $\tilde U$, compared with $\tilde D$, follows from the larger electric charge.  Note that the above limits correspond to the lightest  scalar components of $\tilde U$ and $\tilde D$; because these fields are vector-like, there is a complex scalar to begin with that is generically split by the $b$-term $b_U \tilde U \tilde{\bar U} + \mathrm{h.c.}$ (and similarly for $\tilde D$). 

At hadron colliders, there are two possibly relevant search channels: inclusive dijet and dijet pairs.  Unfortunately, searches for inclusive dijets face enormous QCD backgrounds, and the limits from Tevatron and the LHC are 3-4 orders of magnitude too weak to constrain the production of $\tilde U / \tilde D$.  A more promising final state is two pairs of dijets in the same event, with the same mass.  This channel has been pursued, for example, by ATLAS with 34~pb$^{-1}$~\cite{Aad:2011yh} and CMS with 2.2~fb$^{-1}$\cite{CMS-PAS-EXO-11-016}.   ATLAS constrains masses in the 100-200 GeV range and sets a limit of 1000 pb at 100 GeV, which is about a factor of 3 too weak to constrain $\tilde U/ \tilde D$ production~\cite{Beenakker:2010nq}.  CMS constrains masses in the 300-1200 GeV range and sets a limit of 1 pb at 300 GeV, which roughly coincides with the cross-section for $\tilde U / \tilde D$ at this mass.  For heavier masses, the limit is too weak.   Therefore, there is no LHC limit on $\tilde U / \tilde D$ presently (except possibly for a very narrow range of masses near 300 GeV), but the existing searches for two pairs of dijets are at the edge of discovering or constraining these states.

It is also possible to constrain $\tilde D$ through top decays, when $m_{\tilde D} < m_t$.  In particular the flavor combination $\lambda_D^{3i}$ leads to the decay $t \rightarrow \tilde D d_i$, which contributes extra hadronic top decays.  The hadronic top cross-section, and overall top width, are both poorly constrained.  The strongest limit on $\lambda_D^{3i}$  actually comes from the dileptonic top production cross-section, which has been measured by ATLAS to better than 10\% precision~\cite{ATLAS-CONF-2012-024}.  A partial width of top decays into $\tilde D$ depletes the leptonic branching fraction and therefore the dileptonic production cross-section.  We find the constraint $\lambda_D^{3i} < 0.27$ in the limit  $m_{\tilde D} \ll m_t$.

\subsection{Trijets in CRPV}
\label{sec:trijet}

The fermionic $D$ and $U$ states each decay into three-jet resonances.  $D$ ($U$) decays to a jet and an on or off-shell $\tilde U$ ($\tilde D$), which decays to two jets.  If light enough, these states would have been produced at LEP2, resulting in 6 jets.  This final state is constrained by LEP2 searches for the pair production of neutralinos or charginos that decay to three jets through conventional RPV\@.  The L3 RPV search~\cite{Achard:2001ek} presents the cross-section limit versus resonance mass.  By comparing to the leading order cross-section for $D/U$ from Madgraph~\cite{Alwall:2011uj}, we find the limits $m_D > 72$~GeV and $m_U > 85$~GeV at 95\% C.L.  Note that these limits correspond to the three-body decay where $D/U$ decays through an off-shell $\tilde U^*/ \tilde D^*$, since the chargino/neutralino signal used by L3 consists of 3-body decays through off-shell squarks.

The pair production of three-jet resonances has also been searched for by CDF~\cite{Aaltonen:2011sg} and CMS~\cite{Chatrchyan:2011cj, CMStrijet5}.  These searches attempt to reconstruct the 3-jet mass.  In order to solve the combinatoric problem (there are 20 ways to pair 6 jets into two groups of 3), the CDF and CMS searches select triplets of jets with high $\Sigma p_T$.  CDF searches for three jet resonances with mass between 77 and 240 GeV.  By comparing the CDF cross-section limit to the NNLO production cross-section for a heavy quark~\cite{Aliev:2010zk}, we find that CDF excludes  $D/U$ between 77 and 90 GeV at 95\% C.L.  This extends the $L3$ limit, resulting in the combined limit $m_{D/U} > 90$~GeV when $D/U$ experience a three-body decay.  There have been two CMS searches conducted at $\sqrt s = 7$~TeV: the first search used a luminosity of 30 pb$^{-1}$ and constraints trijet resonances with masses between 200 and 500 GeV.  The second search used 5 fb$^{-1}$ and constraints trijet resonances between 300-1600 GeV.  Neither of these searches has the sensitivity to constrain $U/D$.  For example, the 35 pb$^{-1}$ search sets a limit at 200 GeV of 400 pb, compared to the $U/D$ production cross-section of 77 pb.

Although the CMS searches for three-jet resonances do not have the sensitivity to constrain $D/U$ directly, they can constrain gluino decays that pass through $D$ or $U$, because of the larger gluino cross-section.  The CMS searches were designed to constrain pair production of three-jet resonances, but the search strategies are inclusive enough to capture final states with more jets that contain three jet resonances.  Specifically, in the 35 pb$^{-1}$ (5 fb$^{-1}$) search, events are considered if there are 6 or more jets with $p_T > 45$ (70)~GeV, and events are selected if there exist any combination of three jets with high $\Sigma p_T$.  The three jet mass distribution is then searched for excesses. 

\begin{figure}[!h]
\begin{center} 
\includegraphics[width=\textwidth]{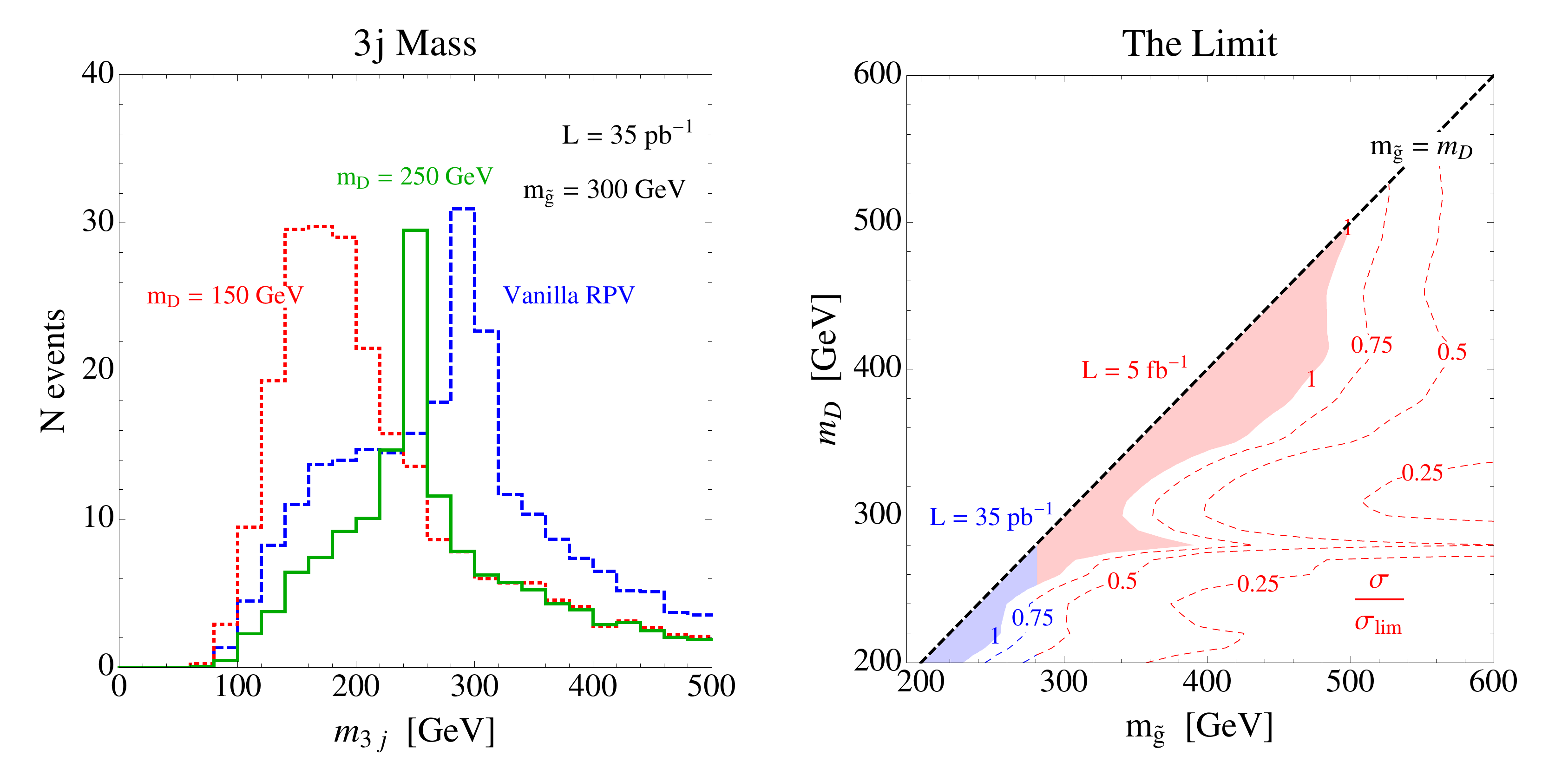} 
\end{center}
\caption{Three jet mass and the limit on it.  To the left, we show the three jet mass after the cuts of the 35 pb$^{-1}$ CMS search for trijet resonances~\cite{Chatrchyan:2011cj}.  We fix the gluino mass to 300 GeV and compare conventional (vanilla) RPV to CRPV with $m_{\tilde U} = 100$~GeV and $m_{D} = 150, 250$~GeV.  We see that the $D$ mass is reconstructed with a similar acceptance and resolution to the gluino mass in conventional RPV.  Assuming the same acceptance, we show the limit on CRPV, to the right, coming from the CMS searches with 35 pb$^{-1}$ and 5 fb$^{-1}$.  The contours show the signal cross-section divided by the cross-section limit.  The shaded red area, where this ratio is larger than 1, is excluded at 95\% C.L.}
\label{fig:3jres}
\end{figure}

 In our model, gluino production results in a 10 jet final state, at parton-level, including a $D$ or $U$ three-jet sub-resonance on each side of the event.  In order to test the acceptance of the CMS searches for our model, we have simulated gluino pair production, and the CRPV cascade, in Pythia~6~\cite{Sjostrand:2006za}, including showering but not hadronization.  We pass the Pythia output through an idealized calorimeter (using $\eta \times \phi$ cells of size $0.09 \times 0.09$) and cluster jets with Fastjet~\cite{Cacciari:2011ma} using the anti-$k_T$ algorithm with $R = 0.5$, as was used by CMS.  After applying the selections of the CMS searches, we find a comparable acceptance for our signal as for gluino production in conventional RPV, which is the signal considered by CMS.  This is shown, for the 35 pb$^{-1}$ search, to the left of figure~\ref{fig:3jres}, where we have fixed $m_{\tilde g} = 300$ GeV, and we show the resulting 3-jet mass for $m_D = 150, 250$~GeV and $m_{\tilde U} = 100$~GeV.  The gluino is assumed to decay to $D$ and two jets, through an off-shell squark.  For comparison, we also show the mass distribution from conventional RPV, where the gluino decays directly to three jets.  We see that the $D$-mass bump is reconstructed with a comparable acceptance and resolution to the gluino mass in conventional RPV.  We find that the same is true for the 5 fb$^{-1}$ search (however, at heavier resonance masses, the reconstructed resolution is degraded for both conventional and CPRV in the new search). This means that the limit on CRPV can be simply inferred from the limit on conventional RPV, as a function of the gluino mass, which sets the signal cross-section, and the $D$ (or $U$) mass, which determines the cross-section limit.  We show the result of this procedure to the right of figure~\ref{fig:3jres}, where, for simplicity, we have assumed that CRPV has the same acceptance as conventional RPV, as motivated by the left of figure~\ref{fig:3jres}.  CRPV is only constrained when $m_{\tilde g} < 300$~GeV and when $m_D$ is close to $m_{\tilde g}$.

\subsection{Quadjets in CRPV}

Our model also contains 4-jet resonances: a signature which has never been searched for in a collider.  For example, squarks decay into a 4-jet final state and squark pair production therefore leads to two 4-jet resonances of the same mass in each event.  Gluino production can also lead to 4-jet resonances downstream, if each gluino decays to on-shell squarks.  It should be possible to constrain these channels if the CDF/CMS search strategy~\cite{Aaltonen:2011sg, Chatrchyan:2011cj} is extended to look for excesses in the 4-jet mass.  As for the 3-jet resonance search, it should be possible to reduce the combinatoric background by selecting combinations of 4-jets with high $\Sigma p_T$.

\begin{figure}[!h]
\begin{center} 
\includegraphics[scale=0.45]{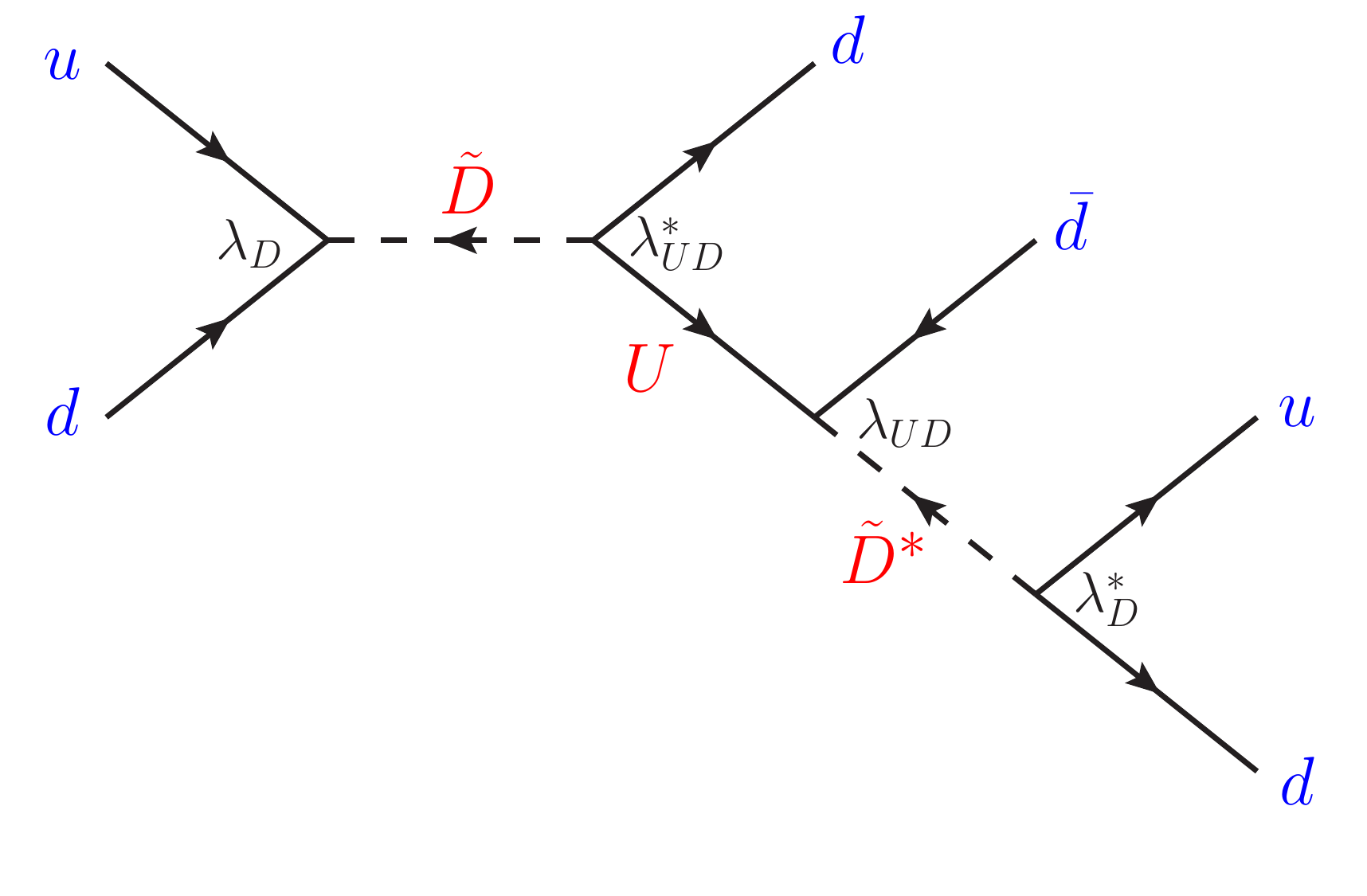} 
\end{center}
\caption{\label{fig:4jFeyn}
A four jet resonance that is possible in CRPV.  This resonance proceeds at a large rate if the couplings $\lambda_D$ and $\lambda_{UD}$ are both large, which is possible without exceeding baryon number violation constraints as long as $\lambda_U$ is sufficiently small.  Unlike conventional RPV, $s$-channel production can proceed through two valence quarks without dangerous baryon number violation.}
\end{figure}

Another exciting possibility in CRPV is the production of a 4-jet resonance in the s-channel, as in figure~\ref{fig:4jFeyn}.  The $\lambda_D$ coupling allows $\tilde D$ to be produced directly in the s-channel.  Squarks can be produced resonantly in conventional RPV scenarios, but not through two valence quarks because the $\lambda''_{11i} u^1 d^1 d^i$ coupling is highly constrained by baryon-number violating processes (for a recent discussion see Ref.~\cite{Kilic:2011sr}).  Therefore, s-channel production of squarks in conventional RPV is necessarily suppressed by non-valence PDFs such as that of the strange quark.  In CRPV, though, the $\lambda_D^{11}$ coupling to first generation quarks does not violate baryon number on its own, because collective breaking is necessary.    If only $\lambda_D^{11}$ is large, the resulting final state is a 2-jet resonance which is constrained by the usual searches for excesses in the dijet spectrum.  Alternatively, if $\lambda_D^{11}$ and $\lambda_{UD}^i$ are both large, there can be a large rate for the 4-jet resonant process depicted in figure~\ref{fig:4jFeyn}, as long as $m_{\tilde D} > m_U$ so that $U$ is produced on-shell.  Note that $\lambda_D$ and $\lambda_{UD}$ can both be large without violating baryon number, if $\lambda_U$ is sufficiently small.  Equivalently, the process of figure~\ref{fig:4jFeyn} alone does not violate baryon number, because we can assign $\tilde D$ baryon number +2/3 and $U$ baryon number -1/3.

\begin{figure}[!h]
\begin{center}
\includegraphics[width=0.7\textwidth]{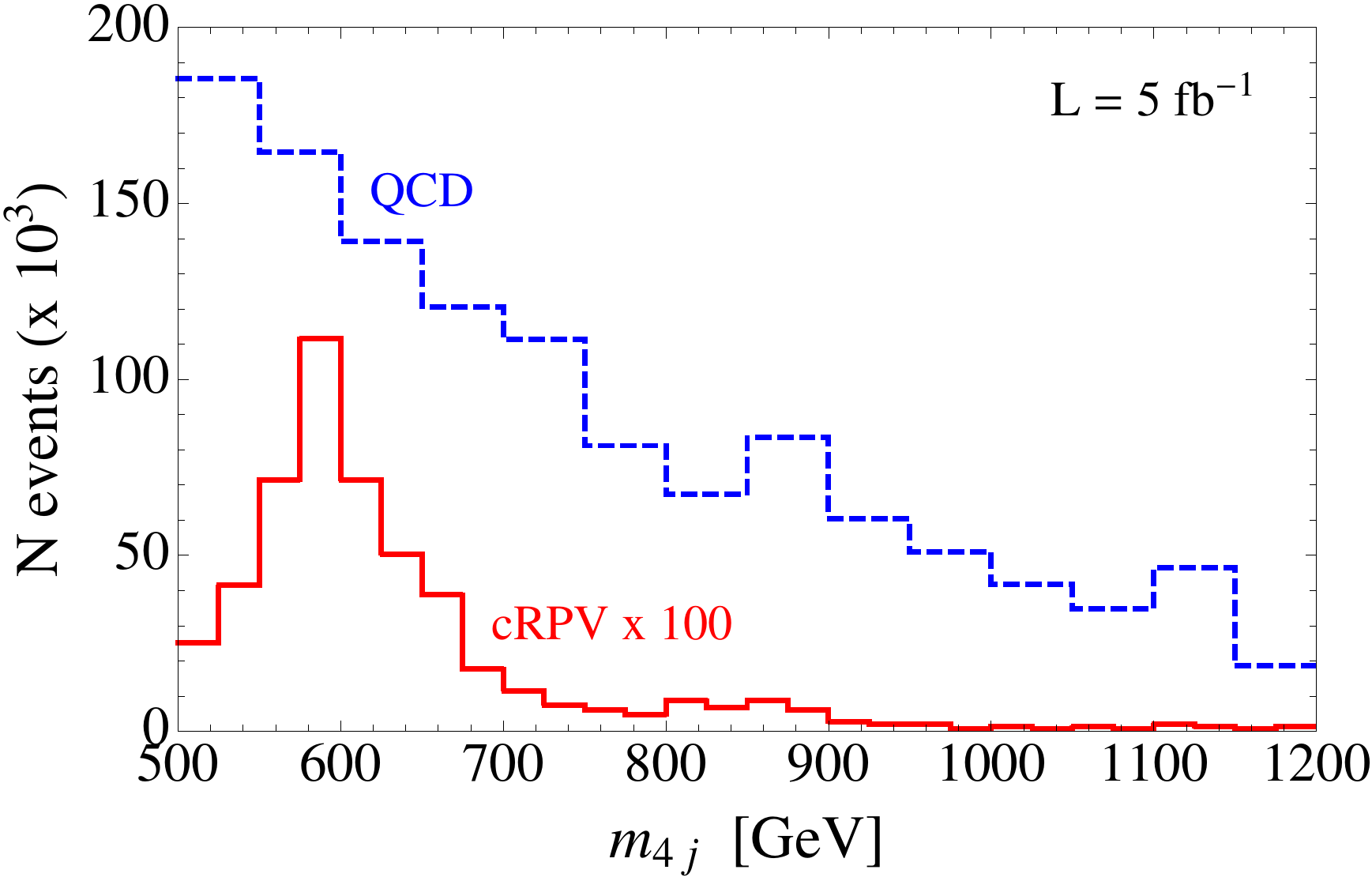}
\end{center}
\caption{\label{fig:4jMass}
The $4j$ mass from CRPV (in red) and QCD (in blue), using the selection described in the text.  The $D$ resonance is clearly visible in the signal, while the QCD background is featureless.  It is possible for the signal significance to exceed $S / \sqrt B$ of 3 in the current data, consistently with other constraints.}
\end{figure}

The CDF/CMS $3j$ strategy cannot be trivially extended to $s$-channel production of 4 jets, because $\tilde D$ will be produced close to threshold without yielding a large $\Sigma p_T$.   We have devised a set of alternate cuts that would allow for the search of an $s$-channel $4j$ resonance.  With a simple MC estimate, we find that it may already be possible to search for such a resonance, above QCD background, in the existing LHC data.  Fixing $m_{\tilde D} = 600$~GeV and $m_{U}=200$~GeV, we have simulated the process of figure~\ref{fig:4jFeyn} using Madgraph~5~\cite{Alwall:2011uj} for the parton-level process and Pythia~6~\cite{Sjostrand:2006za} for showering, but not hadronization.  As in section~\ref{sec:trijet}, the Pythia output is passed through an idealized calorimeter before clustering with Fastjet~\cite{Cacciari:2011ma}, this time using anti-$k_T$ with $R = 0.4$.  We look for events with exactly 4 jets with $p_T > 70$~GeV and $\eta < 3$.  We veto events with a 5th jet of $p_T > 30$~GeV and $\eta < 3$, which helps select for clean signal events and remove events with extra radiation or where one of the parton level jets reconstructs as more than one jet.  We also veto events if the hardest jet has $p_T > 300$~GeV or if the second hardest jet has $p_T > 125$~GeV.  This helps remove signal events where multiple parton-level jets merge during jet reconstruction.  The resulting 4-jet mass distribution is shown in figure~\ref{fig:4jMass}, compared to the QCD background (simulated using Madgraph $2 \rightarrow 3$ plus showering, which gives a larger rate passing our cuts than $2 \rightarrow 2$ or $2 \rightarrow 4$ production).

We see from figure~\ref{fig:4jMass} that the CRPV signal exhibits a clear resonance at the $\tilde D$ mass of 600 GeV, while the QCD background is featureless.  The relevant question to determine discoverability, then, is what signal size is possible, given other constraints.  There are two constraints that limit the signal size: the dijet limit on the process where $\tilde D$ decays to two jets through the $\lambda_D$ coupling, and perturbativity of the $\lambda_{UD}$ coupling.  For example, if we allow  $\lambda_{UD}^1 = 1$, then we find that dijet limits constrain $\lambda_D^{11} < 0.6$~\cite{Aad:2011aj} at 95\% C.L.  The normalization of the signal in figure~\ref{fig:4jMass} is chosen to saturate these values, and we include the finite $\tilde D$ width that corresponds to this choice of couplings.  Within the band 550-650 GeV, we find an LO signal cross-section of 0.6~pb and a QCD cross-section of 140~pb, leading to $S / \sqrt B \sim 3.6$ with 5~fb$^{-1}$.  This illustrates that indeed $s$-channel production of a 4 jet resonance may be observable in the current data.  Admittedly, our QCD background estimation, which is LO and unmatched, is crude.  A realistic background estimate would need to be data-driven: for example experimentalists can fit a power-law to the 4-jet mass and search for excesses, as is done for dijets.

\section{Alternative Models of CRPV}
\label{sec:examples}
The model we have so far discussed has a number of important features: it almost maximally hides SUSY (in all hadronic channels) while simultaneously showing the simple manner in which CRPV can arise. At the same time, it is merely a simple example of the sorts of models that could arise.

The simplest modification to this would be to write down an analogous model, but with leptons. I.e., 
\be
W= L le + llE + LlE.
\ee
Such a model would not, in general, hide SUSY. However, it would lead to dramatic final states of cascades analogous to those in Figure \ref{fig:cdecay}. For instance  $\chi_0 \rightarrow L \tilde L$ with $\tilde L^{-/0} \rightarrow l^{0/+} e^-$ (a dilepton resonance for $L^0$) and $L^{+/0} \rightarrow \tilde E^+ l^{0/-}$, $\tilde E \rightarrow l^+l^0$. 
Alternatively $\chi_0 \rightarrow E \tilde E$ with $\tilde E^+ \rightarrow l^+ l^0$, $E^-  \rightarrow l^- \tilde L^0$ followed by $\tilde L^0 \rightarrow l^- e^+$ (i.e., $E^-$ yields a trilepton resonance).

With, e.g., $\tilde q \tilde q$ production, with $\tilde q \rightarrow q \chi_0$, each neutralino can decay to 2 or 4 leptons +MET. Decays with 4 leptons can contain dilepton or {\em trilepton} resonances. Thus, the overall signal would be dijets + 2,4,6, or 8 leptons, with resonances in the higher lepton multiplicity events.

Such an event would be a clear sign of BSM physics, so it is clearly not something that will hide SUSY. One might argue that requiring high lepton multiplicities is not optimal in these cases because one takes efficiency hits with each additional lepton required. On the other hand, with such high multiplicities, where SM backgrounds are already low, it is at least conceivable that lepton tags with higher fake rates would be acceptable and more optimal search strategies could be found. 

\begin{figure}[t]
\begin{center} \includegraphics[scale=0.55]{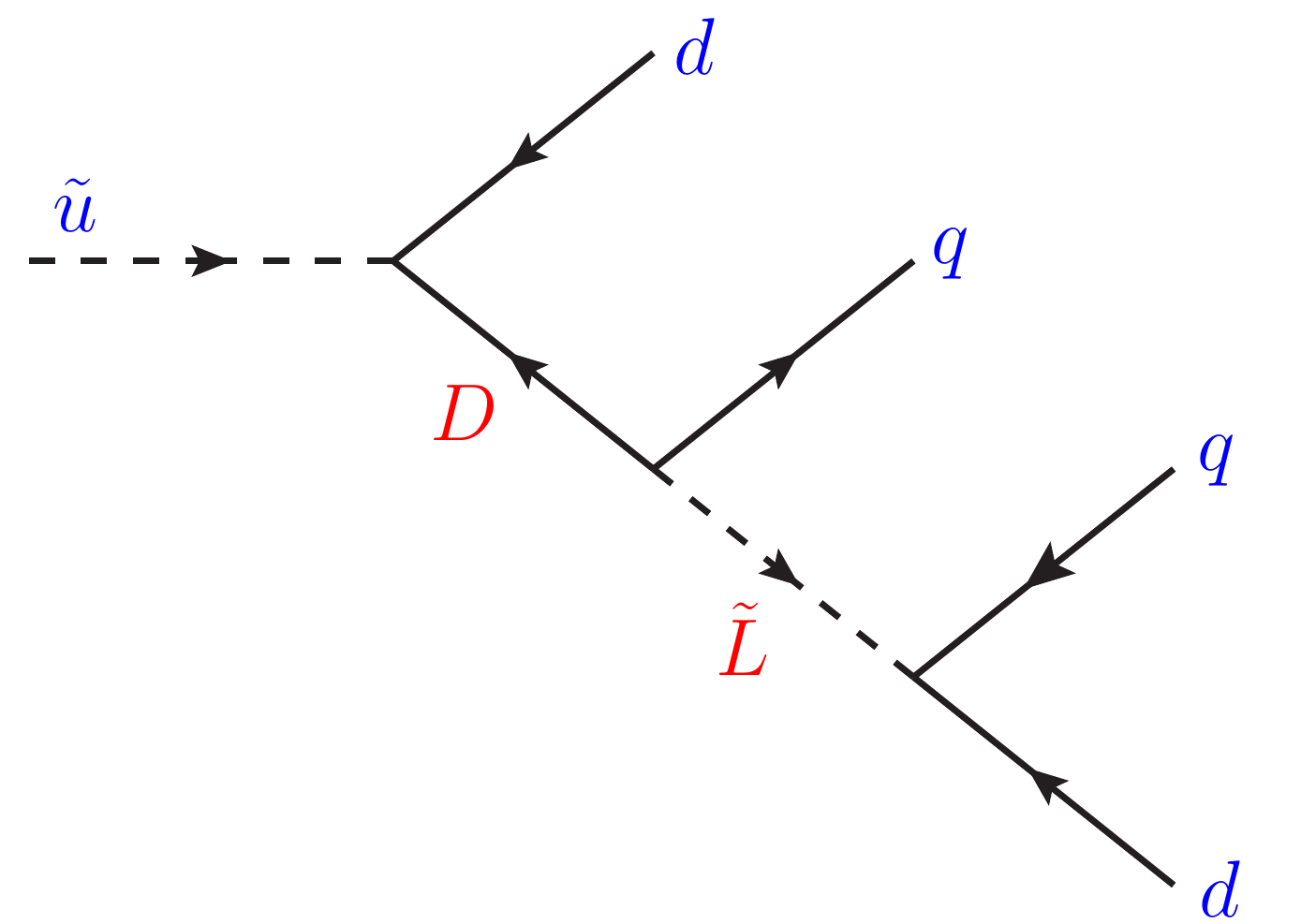} \end{center}
\caption{The decay of the $\tilde u$ in the model of Eq. \ref{eq:55barmodel}.}
\label{fig:55bardecay}
\end{figure}

In the context of GUTs, the models written down so far require both fields from a $\bf 10 + \bar 10$ and a $\bf 5 + \bar 5$. As a consequence, to embed in a GUT a large number additional spectator fields would be needed, and a large number of operators must be ignored. An alternative would be to just try to violate R-parity using {\em only} a $\bf 5 + \bar 5$. A simple such model would be
\be
W = u d D+L q d + L q D.
\label{eq:55barmodel}
\ee
This model preserves lepton number, but violates baryon number collectively. Neutralino decays $\chi_0 \rightarrow \tilde L L$ could be followed by $\tilde L \rightarrow q d$ and $L \rightarrow q \tilde D$ with $\tilde D \rightarrow u d$ (see figure \ref{fig:55bardecay}). Thus, we have the usual $\chi_0 \rightarrow 4 j$. 

We should be clear that while the gauge coupling unification is realized easily here (i.e., the field content will not modify the differential running of the couplings), the proton decay that usually accompanies a triplet Higgs is a potential problem. In particular, we must omit the $q Dl$ coupling by hand or else we are confronted by excessive baryon number violation.

The phenomenology of this model is quite similar to that of the model we have focused on in this paper. However, s-channel $\tilde L$ exchange leads to dangerous contributions to meson mixing, which force smaller couplings in this model than the $UDD$ model of Eq. \ref{eq:crpvmodel}. Nonetheless, it is possibly a simpler UV model than the $UDD$ model if one accepts some additional overall suppression of couplings.

All of these models so far have broken R-parity through a combination of couplings of multiple fields all of which carry SM charges. One can imagine collectively breaking R-parity through a singlet, as well.
Consider
\be
W=\lambda_{Udd} Udd + M_U U \bar U +\lambda_{NUu} N \bar U u +M_N N^2+ \lambda_{NNN} N^3
\label{eq:singletcrpv}
\ee
The first term identifies $U$ as a diquark. The third term then implies that $N$ carries baryon number, which is violated by the $N^3$ term. Aside from the general idea that CRPV could arise with a singlet, this model has a natural extra-dimensional interpretation. Here, the first three terms could exist on a brane, on which the SM fields as well as $U, \bar U$ are confined. $N$ could propagate in the bulk, and a separate brane, where R-parity is explicitly violated could host the $N^3$ term. This gives a natural explanation as to why all other RPV couplings are absent at tree level in this model. 

\begin{figure}[!h]
\begin{center}
\includegraphics[width=0.65\textwidth]{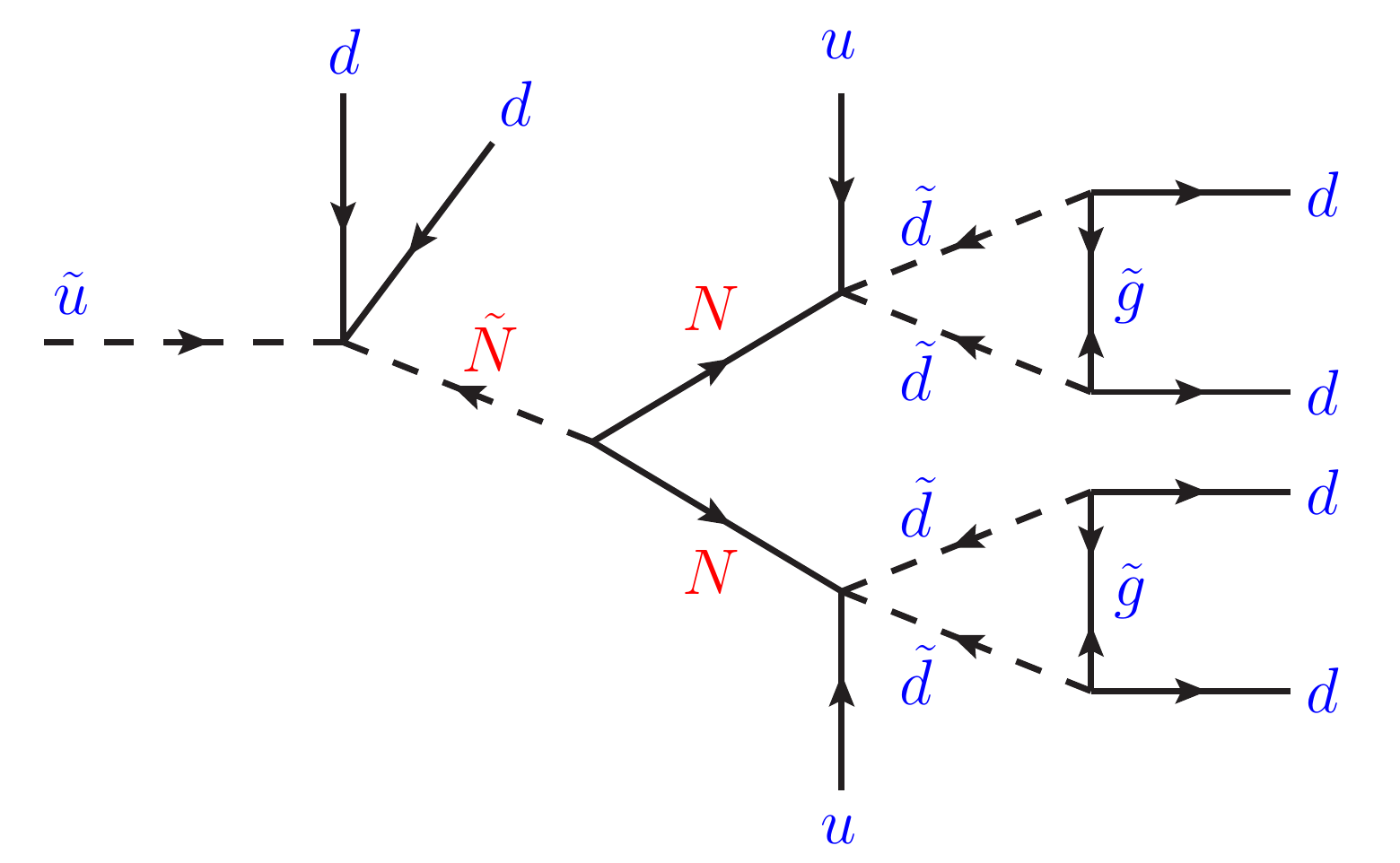}
\end{center}
\caption{\label{fig:Nudddecay}
The cascade decay of $\tilde u$ through intermediate $N$s with the model in Eq. \ref{eq:singletcrpv} and the $U, \bar U$ integrated out.}
\end{figure}

The simplest way to see how a squark would decay in this model is from integrating out the $U, \bar U$. We are left with a higher dimension operator $\lambda_{Udd} \lambda_{NUu} u d d N/M_U$. The production of a $\tilde u$ would be followed by $\tilde u \rightarrow \tilde N dd$. $\tilde N$ can decay through $\lambda_{NNN}$ to $NN$. $N$, in turn, decays through the diagram in Figure \ref{fig:Nudddecay} into three jets (i.e., a neutron). Thus, a single $\tilde u$ decays into a final state of 8 jets, or disquark production decays into 16 jets. (Gluino production results in the only slightly more spectacular 18 jets.)

This resulting dimension 6 decay operator for N is
\be
G_{Nudd} N u dd = \frac{g_{QCD}^2}{16 \pi^2 \Lambda^4}\lambda_{Udd}\lambda_{NUu}N u dd
\ee
where $\Lambda$ is a representative combination scale of $m_{\tilde g}$ and $m_U$,

With this, we can estimate the lifetime of the $N$ to be 
\be
c \tau \approx 1~ {\rm cm} \times \left(\frac{\lambda_{Udd}\lambda_{NUu}}{10^{-4}}\right)^{-2} \left(\frac{m_{\tilde u}}{500 \gev}\right)^{-5} \left(\frac{\Lambda}{{\rm TeV}}\right)^4.
\ee
Thus, not only would such a model yield incredibly high multiplicity jet signatures, they could well be displaced by large amounts from the primary vertex.

This model can be more easily embedded into a GUT. Specifically, we can extend the field content to a complete ${\bf 10} + {\bf \overline{10}}$ and write down
\be
W &\supset& m_Q Q \bar Q+ m_U U \bar U  + m_E E \bar E+ m_N N^2 \nn \\
&+&Q ld +Udd+ llE  \nn \\
&+& N ( q \bar Q + u \bar U+e \bar E) + N^3.
\ee
This model violates B and L, but preserves B-L. A loop diagram through the operators $Q l d, U dd, N q \bar Q, N u \bar U$ will allow proton decay. Altogether, we can recast the limit of \cite{Dreiner:1997uz} to yield 
\be
\lambda_{Qld} \lambda_{Udd}\lambda_{Nq\bar Q} \lambda_{Nu\bar U} \lesssim 10^{-25} \left(\frac{m_{SUSY}}{100 \gev}\right)^2,
\ee
where $m_{SUSY}$ is the characteristic scale of the box. For couplings $\lambda \sim 10^{-6}$ we can satisfy this constraint and still have non-displaced decays. Unlike traditional RPV (where the constraints are $\sim 10^{-13}$ when $udd$ and $lle$ are both present) it is quite easy to have these decays occur within the detector.

Finally, we can consider two models that, in the presence of EWSB, do not look identical to the collective models we have discussed, because they have explicit mass mixing, but nonetheless, look collective in the UV theory. Moreover, they retain the essential features of CRPV in that cascade decays can be prompt, but yield RPV signals ultimately. 

To begin, we can include a vectorlike quark $D_g, \bar D_g$ (elsewhere referred to as ``G-quarks'' \cite{sister2}) as well as the couplings
\be
W=q h_d D_g + \mu D_g \bar D_g + u d D_g.
\ee
Considering squark production $\tilde u$ for instance, if the fermion $D_g$ is lighter than the squark, then $\tilde u \rightarrow D_g d$, with $D_g \rightarrow q H$ would be a natural decay. If $D_g$ is heavier, then once the Higgs acquires a vev, the $u d D$ operator will mediate a conventional RPV-type signal from the $D_g-q$ mixing. I.e., we can integrate out the $D$ fields and are left with a K\"ahler operator $(q h_d)^\dagger d d$. In the presence of a Higgs vev, this allows $\tilde d \rightarrow d q$ similar to the traditional $udd$ decays.

One can try a similar approach with a doublet. I.e., we can include additional vectorlike doublets $\Sigma_u$ and $\Sigma_d$ that acquire vevs (sister Higgs fields), and include the superpotential
\be
W=\Sigma_d h_d e + \mu_h \Sigma_d h_u + \mu \Sigma_d \Sigma_u.
\ee
Such a model is intriguing, in that it induces direct lepton-gaugino mixing {\em without} large contributions to neutrino mass (as with $LH$). On the other hand, it tends to do a poor job of hiding SUSY, in that final states generally contain leptons (and MET in the cases that the leptons are $\tau$'s).

\section{Conclusions}
\label{sec:conclusions}
The absence of any clear sign of supersymmetry so far at the LHC places strong constraints on its properties, if it exists at the weak scale. While one possibility is that the MSSM is simply just out of reach, another possibility is that there is a rich sector of new physics, but in such a way that the new signals do not show clear signs of MET or hard leptons that are classic signatures of SUSY.

A simple explanation for this is the idea that R-parity is not conserved, and that the LSP can decay. Traditional RPV operators are highly constrained, however, not only by B- and L-violating processes, but also by flavor constraints, such that most operators are constrained to be quite small. Indeed, for anarchic RPV operators, the decays arising from it are generally quite displaced, often outside the tracker or even detector at the LHC.

However, in the presence of new fields, and specifically new vectorlike matter fields, new opportunities for RPV arise. In particular, the usual concerns are weakened when R-parity is broken {\em collectively}. When the new fields interact with the SM quarks, their R-parities are unclear initially. Often, it is only with a combination of multiple interactions that R-parity is violated, and no single coupling can be said to break R-parity alone.

This collective R-parity violation has important phenomenological consequences, because the B- and L- violating processes must involve all of the couplings, leading to complicated and suppressed diagrams. Thus, even for fairly large $(\sim 10^{-2})$ couplings, the theory can be safe from these, and other processes.

On the other hand, cascade decays sample only one coupling at a time. Thus, the cascade can be prompt, even while the dangerous processes are suppressed. This can have important consequences for SUSY searches. In particular, in the model we have shown, gluinos could plausibly be as light as 300 GeV and still evade current limits.

Since missing energy is no longer in general present, the new particles produced (i.e., squarks or gluinos) should be completely reconstructable as resonances. However, often the cascades are long enough that the mass is distributed among many particles, and instead the best resonances to reconstruct are lower in the cascade. (I.e., reconstructing a trijet resonance from a $D$ decay produced from a gluino cascade.) Nonetheless, these multi-object resonances (MORs) are ubiquitous in these theories, and while two- and three-jet resonances are presently searched for, these models motivate other, more exotic (detectable) resonances, such as four-jet and three-lepton resonances.

While CRPV may seem strange at first blush, it is in large part simply because it is not present in the MSSM. In models where new vectorlike matter is present, it is not hard to imagine UV completions that would realize it. A simple example is one in which some fields propagate in an extra dimension and interact with a brane where either a different or no R-parity is preserved. In such models, traditional RPV operators would be absent, but these new operators would be present.

There is a tendency to look at the results of the LHC and interpret them as signs that there is nothing new at low energies, but we must be mindful of the limitations of these reactions. If new colored states are present at the LHC, it is entirely plausible that our expected signals of MET could be transferred into signals of multijets where signs of new physics are more challenging. As the LHC progresses, if no signs of MET are to be had, scenarios such as CRPV motivate a reexamination of our assumptions about what sorts of signals may lie, and more importantly may be found, in hadronic channels.

\begin{acknowledgments}
We thank Spencer Chang, Clifford Cheung, Richard Gipstein and Michele Papucci for helpful conversations.  We also thank the KITP for its hospitality while this work was initiated.  NW is supported by NSF grant \#0947827. J.T.R. is supported by a fellowship from the Miller Institute for Basic Research in Science. TRS is supported by NSF grants PHY-0969448 and AST-0807444.
\end{acknowledgments}

\bibliographystyle{apsrev}

\begin{thebibliography}{99}

  \bibitem{ATLASSUSY} 
ATLAS Collaboration, ATLAS Supersymmetry (SUSY) searches, \\
\href{https://twiki.cern.ch/twiki/bin/view/AtlasPublic/SupersymmetryPublicResults}{https://twiki.cern.ch/twiki/bin/view/AtlasPublic/SupersymmetryPublicResults}

\bibitem{CMSSUSY}
CMS Collaboration, CMS Supersymmetry Physics Results, \\
\href{https://twiki.cern.ch/twiki/bin/view/CMSPublic/PhysicsResultsSUS}{https://twiki.cern.ch/twiki/bin/view/CMSPublic/PhysicsResultsSUS}

\bibitem{Fan:2011yu} 
  J.~Fan, M.~Reece and J.~T.~Ruderman,
  JHEP {\bf 1111}, 012 (2011)
  [arXiv:1105.5135 [hep-ph]].
  
\bibitem{Fan:2012jf} 
  J.~Fan, M.~Reece and J.~T.~Ruderman,
  arXiv:1201.4875 [hep-ph].
  
\bibitem{LeCompte:2011fh} 
  T.~J.~LeCompte and S.~P.~Martin,
  Phys.\ Rev.\ D {\bf 85}, 035023 (2012)
  [arXiv:1111.6897 [hep-ph]].
  
\bibitem{Murayama:2012jh} 
  H.~Murayama, Y.~Nomura, S.~Shirai and K.~Tobioka,
  arXiv:1206.4993 [hep-ph].
  
\bibitem{Baryakhtar:2012rz} 
  M.~Baryakhtar, N.~Craig and K.~Van Tilburg,
  arXiv:1206.0751 [hep-ph].

\bibitem{Hall:1983id} 
  L.~J.~Hall and M.~Suzuki,
  Nucl.\ Phys.\ B {\bf 231}, 419 (1984).

\bibitem{Dreiner:1997uz}
  H.~K.~Dreiner,
  arXiv:hep-ph/9707435.
    
\bibitem{Barbier:2004ez}
  R.~Barbier {\it et al.},
  Phys.\ Rept.\  {\bf 420}, 1 (2005)
  [arXiv:hep-ph/0406039].

\bibitem{Graham:2012th} 
  P.~W.~Graham, D.~E.~Kaplan, S.~Rajendran and P.~Saraswat,
  arXiv:1204.6038 [hep-ph].
  
\bibitem{Brust:2012uf} 
  C.~Brust, A.~Katz and R.~Sundrum,
  arXiv:1206.2353 [hep-ph].
  
\bibitem{D'Ambrosio:2002ex} 
  G.~D'Ambrosio, G.~F.~Giudice, G.~Isidori and A.~Strumia,
  Nucl.\ Phys.\ B {\bf 645}, 155 (2002)
  [hep-ph/0207036].
  
\bibitem{Nikolidakis:2007fc} 
  E.~Nikolidakis and C.~Smith,
  Phys.\ Rev.\ D {\bf 77}, 015021 (2008)
  [arXiv:0710.3129 [hep-ph]].
  
\bibitem{Smith:2008ju} 
  C.~Smith,
  arXiv:0809.3152 [hep-ph].
  
\bibitem{Csaki:2011ge} 
  C.~Csaki, Y.~Grossman and B.~Heidenreich,
  Phys.\ Rev.\ D {\bf 85}, 095009 (2012)
  [arXiv:1111.1239 [hep-ph]].

\bibitem{ArkaniHamed:2001nc} 
  N.~Arkani-Hamed, A.~G.~Cohen and H.~Georgi,
  Phys.\ Lett.\ B {\bf 513}, 232 (2001)
  [hep-ph/0105239].

\bibitem{ArkaniHamed:2002qx} 
  N.~Arkani-Hamed, A.~G.~Cohen, E.~Katz, A.~E.~Nelson, T.~Gregoire and J.~G.~Wacker,
  JHEP {\bf 0208}, 021 (2002)
  [hep-ph/0206020].
  
\bibitem{ArkaniHamed:2002qy} 
  N.~Arkani-Hamed, A.~G.~Cohen, E.~Katz and A.~E.~Nelson,
  JHEP {\bf 0207}, 034 (2002)
  [hep-ph/0206021].

  \bibitem{ATLASHiggs}
  ATLAS Collaboration,
ATLAS-CONF-2012-093, July 2012.

  \bibitem{CMSHiggs}
  CMS Collaboration,
CMS-PAS-HIG-12-020, July 2012.

\bibitem{Hall:2011aa} 
  L.~J.~Hall, D.~Pinner and J.~T.~Ruderman,
  JHEP {\bf 1204}, 131 (2012)
  [arXiv:1112.2703 [hep-ph]].
  
\bibitem{Batra:2003nj} 
  P.~Batra, A.~Delgado, D.~E.~Kaplan and T.~M.~P.~Tait,
  JHEP {\bf 0402}, 043 (2004)
  [hep-ph/0309149].
  
\bibitem{Maloney:2004rc} 
  A.~Maloney, A.~Pierce and J.~G.~Wacker,
  JHEP {\bf 0606}, 034 (2006)
  [hep-ph/0409127].
  

\bibitem{sister1} 
  D.~S.~M~Alves and P.~J.~Fox and Shih, N.~J.~Weiner,
  arXiv:1207.5499 [hep-ph].
  
  \bibitem{sister2} 
  D.~S.~M~Alves and P.~J.~Fox and Shih, N.~J.~Weiner,
  arXiv:1207.xxxx [hep-ph].
  
\bibitem{Goity:1994dq} 
  J.~L.~Goity and M.~Sher,
  Phys.\ Lett.\ B {\bf 346}, 69 (1995)
  [Erratum-ibid.\ B {\bf 385}, 500 (1996)]
  [hep-ph/9412208].
  
\bibitem{Choi:1996nk}
  K.~Choi, E.~J.~Chun and J.~S.~Lee,
  Phys.\ Rev.\  D {\bf 55}, 3924 (1997)
  [arXiv:hep-ph/9611285].

  \bibitem{pdg}
J. ~Beringer et al. (Particle Data Group), PR D86, 010001 (2012) (URL: http://pdg.lbl.gov)

\bibitem{Barbieri:1985ty} 
  R.~Barbieri and A.~Masiero,
  Nucl.\ Phys.\ B {\bf 267}, 679 (1986).

\bibitem{Isidori:2010kg} 
  G.~Isidori, Y.~Nir and G.~Perez,
  Ann.\ Rev.\ Nucl.\ Part.\ Sci.\  {\bf 60}, 355 (2010)
  [arXiv:1002.0900 [hep-ph]].

\bibitem{Dreiner:1992vm} 
  H.~K.~Dreiner and G.~G.~Ross,
  Nucl.\ Phys.\ B {\bf 410}, 188 (1993)
  [hep-ph/9207221].
  
\bibitem{Davidson:1998za} 
  S.~Davidson,
  In *Trento 1998, Lepton and baryon number violation* 394-414
  [hep-ph/9808427].
  
\bibitem{Endo:2009cv} 
  M.~Endo, K.~Hamaguchi and S.~Iwamoto,
  JCAP {\bf 1002}, 032 (2010)
  [arXiv:0912.0585 [hep-ph]].
  
\bibitem{Heister:2002jc} 
  A.~Heister {\it et al.}  [ALEPH Collaboration],
  Eur.\ Phys.\ J.\ C {\bf 31}, 1 (2003)
  [hep-ex/0210014].
  
\bibitem{Aaltonen:2011sg}
  T.~Aaltonen {\it et al.}  [CDF Collaboration],
  arXiv:1105.2815 [hep-ex].

\bibitem{Chatrchyan:2011cj} 
  S.~Chatrchyan {\it et al.}  [CMS Collaboration],
  Phys.\ Rev.\ Lett.\  {\bf 107}, 101801 (2011)
  [arXiv:1107.3084 [hep-ex]].
 
 \bibitem{CMStrijet5}
 CMS Collaboration, ``Updated Search for Three-Jet Resonances in pp Collisions at $\sqrt s =7$~TeV,"
 \href{https://twiki.cern.ch/twiki/bin/view/CMSPublic/PhysicsResultsEXO11060}
 {https://twiki.cern.ch/twiki/bin/view/CMSPublic/PhysicsResultsEXO11060}
  
\bibitem{Aad:2011yh} 
  G.~Aad {\it et al.}  [ATLAS Collaboration],
  Eur.\ Phys.\ J.\ C {\bf 71}, 1828 (2011)
  [arXiv:1110.2693 [hep-ex]].
  
  \bibitem{CMS-PAS-EXO-11-016}
  CMS Collaboration,
CMS-PAS-EXO-11-016, January 2012.

\bibitem{Beenakker:2010nq} 
  W.~Beenakker, S.~Brensing, M.~Kramer, A.~Kulesza, E.~Laenen and I.~Niessen,
  JHEP {\bf 1008}, 098 (2010)
  [arXiv:1006.4771 [hep-ph]].

  \bibitem{ATLAS-CONF-2012-024}
  ATLAS Collaboration,
ATLAS-CONF-2012-024, March 2012.
  
\bibitem{Achard:2001ek} 
  P.~Achard {\it et al.}  [L3 Collaboration],
  Phys.\ Lett.\ B {\bf 524}, 65 (2002)
  [hep-ex/0110057].
  
\bibitem{Alwall:2011uj} 
  J.~Alwall, M.~Herquet, F.~Maltoni, O.~Mattelaer and T.~Stelzer,
  JHEP {\bf 1106}, 128 (2011)
  [arXiv:1106.0522 [hep-ph]].

 
\bibitem{Aliev:2010zk} 
  M.~Aliev, H.~Lacker, U.~Langenfeld, S.~Moch, P.~Uwer and M.~Wiedermann,
  Comput.\ Phys.\ Commun.\  {\bf 182}, 1034 (2011)
  [arXiv:1007.1327 [hep-ph]].
 
\bibitem{Sjostrand:2006za} 
  T.~Sjostrand, S.~Mrenna and P.~Z.~Skands,
  JHEP {\bf 0605}, 026 (2006)
  [hep-ph/0603175].
 
   
\bibitem{Cacciari:2011ma} 
  M.~Cacciari, G.~P.~Salam and G.~Soyez,
  arXiv:1111.6097 [hep-ph].
 
\bibitem{Kilic:2011sr} 
  C.~Kilic and S.~Thomas,
  Phys.\ Rev.\ D {\bf 84}, 055012 (2011)
  [arXiv:1104.1002 [hep-ph]].
 
\bibitem{Aad:2011aj} 
  G.~Aad {\it et al.}  [ATLAS Collaboration],
  New J.\ Phys.\  {\bf 13}, 053044 (2011)
  [arXiv:1103.3864 [hep-ex]].
  
\bibitem{Martin:1997ns} 
  S.~P.~Martin,
  In *Kane, G.L. (ed.): Perspectives on supersymmetry II* 1-153
  [hep-ph/9709356].
  
\end{thebibliography}

\end{document}